\documentclass[apjl]{emulateapj}
\usepackage{comment}
\usepackage{ifthen}
\usepackage{amsmath}
\usepackage{amsfonts}
\usepackage{amssymb}
\usepackage{multirow}
\usepackage{wasysym}
\usepackage{subfigure}
\usepackage{epsfig}

\newcommand{\luna}{{\tt LUNA}}
\newcommand{\multi}{{\sc MultiNest}}
\newcommand{\cofiam}{{\tt CoFiAM}}
\newcommand{\obs}{\mathrm{obs}}
\newcommand{\tru}{\mathrm{true}}

\newboolean{emulateapj}
\setboolean{emulateapj}{true}

\newboolean{astroph}
\setboolean{astroph}{true}

\newboolean{png}
\setboolean{png}{false}

\newboolean{color}
\setboolean{color}{true}

\shortauthors{Kipping et al.}
\shorttitle{IV. The Hunt for Exomoons with Kepler (HEK)}
\ifthenelse{\boolean{emulateapj}}{
    \newcommand{\titledag}{$\dagger$}
}{
    \newcommand{\titledag}{\dagger}
}

\begin{document}

\title {The Hunt for Exomoons with Kepler (HEK):\\
 IV. A Search for Moons around Eight M-Dwarfs
\altaffilmark{\titledag}}

\author{
	{\bf D.~M.~Kipping\altaffilmark{1,2},
             D.~Nesvorn\'y\altaffilmark{3},
             L.~A.~Buchhave\altaffilmark{4,5},\\
             J.~Hartman\altaffilmark{6},
             G.~\'A.~Bakos\altaffilmark{6,7,8},
             A.~R.~Schmitt\altaffilmark{9}
	}
}
\altaffiltext{1}{Harvard-Smithsonian Center for Astrophysics,
		Cambridge, MA 02138, USA; email: dkipping@cfa.harvard.edu}

\altaffiltext{2}{NASA Carl Sagan Fellow}

\altaffiltext{3}{Dept. of Space Studies, Southwest Research Institute, 
1050 Walnut St., Suite 300, Boulder, CO 80302, USA}

\altaffiltext{4}{Niels Bohr Institute, University of Copenhagen, DK-2100, 
		Copenhagen, Denmark}

\altaffiltext{5}{Centre for Star and Planet Formation, Natural History Museum of 
		Denmark, University of Copenhagen, DK-1350, Copenhagen, Denmark}

\altaffiltext{6}{Dept. of Astrophysical Sciences, Princeton University,
		Princeton, NJ 05844, USA}

\altaffiltext{7}{Alfred P. Sloan Fellow}

\altaffiltext{8}{Packard Fellow}

\altaffiltext{9}{Citizen Science}

\altaffiltext{$\dagger$}{
Based on archival data of the \emph{Kepler} telescope. 
}


\begin{abstract}

With their smaller radii and high cosmic abundance, transiting planets around
cool stars hold a unique appeal. As part of our on-going project to measure the 
occurrence rate of extrasolar moons, we here present results from a survey 
focussing on eight \emph{Kepler} planetary candidates associated with M-dwarfs.
Using photodynamical modeling and Bayesian multimodal nested sampling, we find 
no compelling evidence for an exomoon in these eight systems. Upper limits on 
the presence of such bodies probe down to masses of $\sim0.4$\,$M_{\oplus}$ in 
the best case. For KOI-314, we are able to confirm the planetary nature of two 
out of the three known transiting candidates using transit timing variations. Of 
particular interest is KOI-314c, which is found to have a mass of 
$1.0_{-0.3}^{+0.4}$\,$M_{\oplus}$, making it the lowest mass transiting planet 
discovered to date. With a radius of $1.61_{-0.15}^{+0.16}$\,$R_{\oplus}$, 
this Earth-mass world is likely enveloped by a significant gaseous envelope 
comprising $\geq17_{-13}^{+12}$\% of the planet by radius. We also find evidence 
to support the planetary nature of KOI-784 via transit timing, but we advocate 
further observations to verify the signals. In both systems, we infer that the 
inner planet has a higher density than the outer world, which may be indicative 
of photo-evaporation. These results highlight both the ability of
\emph{Kepler} to search for sub-Earth mass moons and the exciting ancillary
science which often results from such efforts.

\end{abstract}



\section{INTRODUCTION}
\label{sec:intro}

In recent years, there has been a concerted effort to determine the occurrence 
rate of small planets around main-sequence stars across a broad range of 
spectral types and orbital period using various observational techniques and 
strategies (e.g. \citealt{howard:2010,mayor:2011,bonfils:2013,fressin:2013,
dressing:2013,dong:2013,petigura:2013}). The emerging consensus from these 
studies is that small planets are indeed common with occurrence rates ranging 
from 5\%-90\% for different size and periods ranges. What remains wholly unclear 
is the occurrence rate of ``large'' ($\gtrsim0.1$\,$M_{\oplus}$) moons around 
viable planet hosts. Like mini-Neptunes and super-Earths, there are no known 
examples of such objects in the Solar System but their prevalence would provide 
deep insights into the formation and evolution of planetary systems, as well as 
providing a potentially frequent seat for life in the cosmos 
\citep{williams:1997,heller:2012,heller:2013,forgan:2013}. 

Determining the occurrence rate of large moons constitutes the primary objective 
the ``Hunt for Exomoons with Kepler'' (HEK) project, which seeks evidence of 
transiting satellites in the \emph{Kepler} data \citep{hek:2012}. By 
establishing the occurrence rate as our primary objective, the strategy of our 
survey requires careful consideration of upper limits and prior inputs and is 
therefore substantially more challenging than a simple ``fishing-trip'' style 
survey. For these reasons, HEK conducts the survey in a rigorous Bayesian 
framework, which comes at the acceptable cost of higher computational demands.

In previous works, we have surveyed eight planetary candidates around G and K 
dwarfs for evidence of exomoons, with no compelling evidence for such an object 
identified thus far \citep{hek:2013,kepler22:2013}. Despite null detections, 
Earth-mass and sub-Earth-mass moons are excluded in many cases \citep{hek:2013}. 
Although there have been no transiting exomoon candidates published at this 
time, recently \citet{bennett:2013} reported a candidate free floating 
planet-moon pair via microlensing for MOA-2011-BLG-262. Unfortunately, this 
object cannot be distinguished from a high velocity planetary system in the 
Galactic bulge \citep{bennett:2013}, nor is there much prospect of obtaining a 
repeat measurement to confirm the signal. At this time, the sample of planets 
for which statistically robust limits has been determined is too small to broach 
the question of occurrence rates. The purpose of this work is to extend the 
sample of planetary candidates which have been systematically and 
thoroughly examined for evidence of exomoons.

For our second systematic survey, we focus on planetary candidates orbiting
M-dwarfs ($T_{\mathrm{eff}}<4000$\,K) in the \emph{Kepler} sample. Although 
considerably rarer than FGK hosts in a magnitude limited survey like 
\emph{Kepler} \citep{dressing:2013}, M-dwarfs offer several major advantages for 
seeking exomoons. The most crucial advantage is that since the stars are 
typically two to three times smaller than a Sun-like host, an Earth-sized moon 
produces a four to nine times greater transit depth. This advantage is somewhat
tempered by the lower intrinsic luminosities of the targets, leading to
fainter apparent magnitudes (median \emph{Kepler} magnitude of our sample is 
14.8) and thus higher photometric noise. However, at such faint magnitudes, 
\emph{Kepler's} noise budget is photon-dominated and thus the chance of 
instrumental and time-correlated noise inducing spurious photometric signals 
masquerading as moons is considerably attenuated. Finally, we note that for all 
things being equal except the spectral type of a host star, a planet's Hill 
radius for stable moon orbits is modestly larger (by 25-45\%) in M-dwarf 
systems. The great unknown we cannot factor into our choice of spectral types is 
whether the underlying occurrence rate of large moons is fundamentally distinct 
for M-dwarfs than other spectral types, since no confirmed detections of 
exomoons exist at this time.

\section{METHODS}
\label{sec:methods}


\subsection{Target Selection (TS)}
\label{sub:TS}

From the thousands KOIs (Kepler Objects of Interest) known at this time, we
aim to trim the sample down to just eight objects for our survey. This is
necessary since each object requires decades worth of computational time to 
process \citep{kepler22:2013}. The first cut we make was defined earlier in the
introduction, where we only consider host stars for which 
$T_{\mathrm{eff}}<4000$\,K. Reliable stellar parameters are challenging to
determine for these cool faint stars but recent near-infrared spectroscopy
campaigns by \citet{muirhead:2012} and \citet{muirhead:2014} provide
arguably the most accurate estimates for the cool KOIs. We therefore limit our
sample to only M-dwarfs included in these two catalogs, which contains
203 KOIs spread over 134 host stars. 

In this work, Target Selection (TS) is conducted using the Target Selection
Automatic (TSA) algorithm described in \citet{hek:2012} and updated in 
\citet{hek:2013}. From the 203 KOIs in the \citet{muirhead:2012} and 
\citet{muirhead:2014} catalogs, we only further consider KOIs which satisfy the 
following criteria: i) dynamically capable of maintaining a retrograde 
Earth-mass moon for 5\,Gyr following the calculation method outlined in 
\citet{hek:2012} (which uses the expressions of \citealt{barnes:2002} and 
\citealt{domingos:2006}), ii) an Earth-sized transit with the same duration as 
that of the KOI would be detectable to $\Sigma\mathrm{SNR}>7.1$ when all transit 
epochs are used and iii) an Earth-sized transit with the same duration as that 
of the KOI would be detectable to $\mathrm{SNR}_i>1$ for a single transit epoch. 
These three constraints are the same as those in \citet{hek:2012} except that we 
slightly modify the SNR equation to account for transit duration more 
appropriately using

\begin{align}
\mathrm{SNR}_i &= \frac{ (R_{\oplus}/R_{\star})^2 }{ \mathrm{CDPP}_6 } \sqrt{\frac{T_{14}}{6\,\mathrm{hours}}},\\
\Sigma\mathrm{SNR} & = \mathrm{SNR}_i \sqrt{N_{\mathrm{transits}}},
\label{eqn:SNR}
\end{align}

where CDPP$_6$ is the Combined Differential Photometric Precision over 6 hours
\citep{christiansen:2012}, $T_{14}$ is the first-to-fourth contact duration,
$N_{\mathrm{transits}}$ is the number of usable transits in the time series,
$R_{\oplus}$ is one Earth-radius and $R_{\star}$ is the stellar radius.
These filters leave us with 27 KOIs from which we have selected eight
objects to study in this work. The eight planetary candidates are listed
along with some basic parameters in Table~\ref{tab:planets}. Additionally,
we list the employed stellar parameters of each host star in 
Table~\ref{tab:stars}.

We noted that KOI-314, being an unusually bright \emph{Kepler} M-dwarf at 12.9 
apparent magnitude, has considerably more follow-up available than the other 
seven targets. Specifically, \citet{pineda:2013} have derived stellar parameters 
for this object by stacking high-resolution spectra taken from Keck/HIRES and 
comparing to empirically derived templates of well-characterzied M-dwarfs. We 
therefore use the \citet{pineda:2013} stellar parameters rather than those of
\citet{muirhead:2014} for this unique case. \citet{pineda:2013} do not provide
the effective temperature, however, and so we defer to the measurement of
\citet{mann:2013} for this parameter.

\begin{table*}
\caption{\emph{General properties of the planet candidates studied in this work
and used as inputs for our TSA algorithm. SNR (Signal-to-Noise Ratio) is 
defined in Equation~\ref{eqn:SNR}. Planetary radii were taken from the 
\citet{muirhead:2012}, except for KOI-1596.02 for which we use 
\citet{dressing:2013} radius.
}} 
\centering 
\begin{tabular}{l c c c c c c c c} 
\hline
KOI & \vline & $P_P$ [days] & $R_P$ [$R_{\oplus}$] & $\mathrm{SNR}_i$ & $\bar{\mathrm{SNR}}$ & Multiplicity & $T_{\mathrm{eq}}$ [K] & $K_P$ \\ [0.5ex] 
\hline
KOI-463.01  & \vline &  18.5 & 1.50 & 3.27 & 30.37 & 1 & 268 & 14.7 \\ 
KOI-314.02  & \vline &  23.0 & 1.68 & 4.07 & 33.74 & 3 & 431 & 12.9 \\ 
KOI-784.01  & \vline &  19.3 & 1.67 & 1.20 & 10.86 & 2 & 395 & 15.4 \\ 
KOI-3284.01  & \vline &  35.2 & 0.91 & 2.24 & 15.02 & 1 & 286 & 14.5 \\ 
KOI-663.02  & \vline &  20.3 & 1.58 & 3.52 & 31.16 & 2 & 534 & 13.5 \\ 
KOI-1596.02  & \vline & 105.4 & 2.35 & 1.87 &  7.27 & 2 & 308 & 15.2 \\ 
KOI-494.01  & \vline &  25.7 & 1.65 & 2.23 & 17.50 & 1 & 405 & 14.9 \\ 
KOI-252.01  & \vline &  17.6 & 2.36 & 1.43 & 13.59 & 1 & 472 & 15.6 \\ [1ex] 
\hline\hline 
\end{tabular}
\label{tab:planets} 
\end{table*}

\begin{table*}
\caption{\emph{General properties of the planet candidate host stars studied in 
this work. Last column provides the reference from which we lift the stellar
parameters. $\dagger$ implies effective temperature comes from 
\citet{mann:2013} instead of the quoted reference.
}} 
\centering 
\begin{tabular}{l c c c c c c} 
\hline
KOI & $T_{\mathrm{eff}}$\,[$K$] & $\log g$ & $M_*$ [$M_{\odot}$] & $R_*$ [$R_{\odot}$] & Sp. Type & Reference \\ [0.5ex] 
\hline
KOI-463 & $3389_{-41}^{+57}$ & $4.96_{-0.13}^{+0.14}$ & $0.32\pm0.05$ & $0.31\pm0.04$ & M3V & \citet{muirhead:2014} \\ 
KOI-314 & $3871_{-58}^{+58}$ $^{\dagger}$ & $4.73_{-0.09}^{+0.09}$ & $0.57\pm0.05$ & $0.54\pm0.05$  & M1V & \citet{pineda:2013} \\ 
KOI-784 & $3767_{-51}^{+135}$ & $4.78_{-0.12}^{+0.13}$ & $0.51\pm0.06$ & $0.48\pm0.06$ & M1V & \citet{muirhead:2012} \\ 
KOI-3284 & $3748_{-100}^{+50}$ & $4.75_{-0.07}^{+0.08}$ & $0.55\pm0.04$ & $0.52\pm0.04$ & M1V & \citet{muirhead:2014} \\ 
KOI-663 & $3834_{-57}^{+50}$ & $4.78_{-0.12}^{+0.13}$ & $0.51\pm0.06$ & $0.48\pm0.06$ & M1V & \citet{muirhead:2012} \\ 
KOI-1596 & $3880_{-103}^{+143}$ & $4.77_{-0.14}^{+0.14}$ & $0.51\pm0.07$ & $0.49\pm0.07$ & M0V & \citet{muirhead:2012} \\ 
KOI-494 & $3789_{-159}^{+219}$ & $4.78_{-0.20}^{+0.22}$ & $0.50\pm0.10$ & $0.48\pm0.10$ & M1V & \citet{muirhead:2012} \\ 
KOI-252 & $3745_{-71}^{+53}$ & $4.76_{-0.06}^{+0.06}$ & $0.54\pm0.03$ & $0.51\pm0.03$ & M1V & \citet{muirhead:2014} \\ [1ex] 
\hline\hline 
\end{tabular}
\label{tab:stars} 
\end{table*}

\subsection{Detrending with \cofiam}
\label{sub:detrending}

In order to search for the very small expected amplitudes of exomoon signals
and accurately model the planetary transits, it is necessary to remove the
flux variations due to both instrumental effects and several astrophysical 
effects. In what follows, we use the Simple Aperture Photometry (SAP)
for quarters 1-15 and perform our own detrending rather than relying on the 
Presearch Data Conditioning (PDC) data. Short-cadence (SC) data is used 
preferentially to long-cadence (LC) data wherever available. Short-term 
variations such as flaring, pointing tweaks and safe-mode recoveries (charge 
trapping) are removed manually by simple clipping techniques. Long-term 
variations, such as focus drift and rotational modulations, are detrended using 
the Cosine Filtering with Autocorrelation Minimization (\cofiam) algorithm. 
\cofiam\ was specifically developed to aid in searching for exomoons and we 
direct the reader to our previous papers \citet{hek:2012,hek:2013} for a 
detailed description.

To summarize the key features of \cofiam, it is essentially a Fourier-based
method which removes periodicities occurring at timescales greater than the
known transit duration. This process ensures that the transit profile is
not distorted and if we assume that a putative exomoon does not display a
transit longer than the known KOI, then the moon's signal is also guaranteed to 
be protected. \cofiam\ does not directly attempt to remove high frequency noise 
since this process could also easily end up removing the very small moon signals
we seek. However, \cofiam\ is able to attempt dozens of different harmonics
and evaluate the autocorrelation at a pre-selected timescale (we use 
30\,minutes) and then select the harmonic order which minimizes this 
autocorrelation, as quantified using the Durbin-Watson statistic. This 
``Autocorrelation Minimization'' component of \cofiam\ provides optimized data
for subsequent analysis. In contrast to the initial sample studied in 
\citet{hek:2013}, we found that none of the eight detrended KOI light curves
retained significant ($>3$\,$\sigma$) autocorrelation after \cofiam, which is
consistent with a photon-noise dominated sample, which one might expect for
these generally fainter targets.

\subsection{Light Curve Fits}
\label{sub:fits}

The transit light curves for a planet-with-moon scenario (model $\mathcal{S}$) 
are modeled using the analytic photodynamical algorithm \luna\ 
\citep{luna:2011}, as with previous HEK papers. Photodynamical modeling accounts 
for not only the transits of the planet and the moon but also the transit timing 
and duration variations (TTVs and TDVs) expected \citep{kipping:2009a,
kipping:2009b} in a dynamic framework. These fits always assume just a single
moon and thus our search is implicitly limited to such cases. For comparison, we 
consider a simple planet-only model (model $\mathcal{P}$) as well, for which we 
employ the standard \citet{mandel:2002} routine. Finally, we perform a planet 
fit on each individual transit to derive TTVs and TDVs (model $\mathcal{I}$), 
which is useful in comparing our moon model against perturbing planet models 
later on.

Light curve fits are performed in a Bayesian framework with the goal of both
deriving parameter posteriors and computing the Bayesian evidence 
($\mathcal{Z}$) for each model. The moon fits are particularly challenging 
requiring 14 free parameters exhibiting a large number of modes and complex 
inter-parameter correlations. To this end, we employ the multimodal nested 
sampling algorithm \multi\ \citep{feroz:2008,feroz:2009} as with previous HEK 
papers. For all fits, we use 4000 live points with a target efficiency of 0.1
and use the same parameter sets and priors described in \citet{hek:2013},
giving us 7 free parameters for model $\mathcal{P}$ and 14 for $\mathcal{S}$.
The only change from the \citet{hek:2013} priors, is that we fit the quadratic 
limb darkening coefficients using the $q_1$-$q_2$ parameter set suggested in
\citet{LDfitting:2013}, which provides more efficient parameter exploration.
Contamination factors for each quarter are accounted for using the method
devised in \citet{kiptin:2010} and long-cadence data is resampled
using $N_{\mathrm{resam}}=30$ following the technique devised in 
\citet{binning:2010}.

In \citet{hek:2013}, we described four basic detection criteria (B1-B4) to test
whether an exomoon fit can be further considered as a candidate or not. The 
criteria essentially demand that the moon signal is both significant and
physically reasonable:

\begin{itemize}
\item[{\textbf{B1}}] Improved evidence of the planet-with-moon fits at 
$\geq 4$\,$\sigma$ confidence.
\item[{\textbf{B2}}] Planet-with-moon evidences indicate a preference for
a) a non-zero radius moon b) a non-zero mass moon.
\item[{\textbf{B3}}] Parameter posteriors are physical, in particular $\rho_P$.
\item[{\textbf{B4}}] a) Mass and b) radius of the moon converge away from zero.
\end{itemize}

One slight difference to \citet{hek:2013} is that we have split some of the 
basic detection criteria into sub-criteria. This will be useful 
later since it is often easier to evaluate just a subset of the sub-criteria and 
a failure to pass one of these allows us to quickly reject the object as an 
exomoon candidate. Note that in practice we reject any trials for which the 
density of the planet is excessively low and this causes the radius of the moon 
to be non-zero and so criterion B4b is always satisfied by virtue of the priors
used in our model. B4a is defined as being satisfied if there is a less than a 
5\% false-alarm-probability of the satellite-to-planet mass ratio $(M_S/M_P)$ 
being zero using the \citet{lucy:1971} test.

Another change we implement follows the strategy used recently for Kepler-22b 
\citep{kepler22:2013}, where we allow \luna\ to model negative-radius moons 
(which we treat as inverted transits) during model $\mathcal{S}$. By doing so, 
it is no longer necessary to run a separate moon fit to test B2a, where we would 
have previously locked the radius to zero (although this fit is sometimes still 
a useful tool). This trick essentially saves computation time yet retains our 
ability to test sub-criterion B2a. We define that sub-criterion B2a is 
unsatisfied if the 38.15\% quantile (lower 1\,$\sigma$ quantile) of the derived 
satellite-to-planet radius ratio $(R_S/R_P)$ is negative.

Although we use precisely the same definition for B1 as that of previous papers,
we reiterate here that B1 is computed by evaluating if the Bayesian evidences
between models $\mathcal{P}$ and $\mathcal{S}$ indicates a $\geq4$\,$\sigma$
preference for the latter.

If all of the basic detection criteria are satisfied, we also have three 
follow-up criteria described in \citet{hek:2013} to further vet candidates. The 
concept here is to only fit 75\% of the available data in the original transit 
fits and thus deliberately exclude 25\% of time series, which serves as 
``follow-up'' data.

\begin{itemize}
\item[{\textbf{F1}}] All four basic criteria are still satisfied when new data 
is included.
\item[{\textbf{F2}}] The predictive power of the moon model is superior to that 
of a planet-only model.
\item[{\textbf{F3}}] A consistent and statistically enhanced signal is recovered
with the inclusion of more data.
\end{itemize}

The most powerful of these criteria is F2 and this test is performed for all
KOIs studied in this work. In \citet{hek:2013}, the 75\% original data was
considered to be quarters 1-9 and the follow-up data was quarter 10-12. In this
framework, F2 tests the extrapolated best moon model fit. One significant
change in this work is that we now choose the 25\% follow-up data to lie
somewhere around the middle of our total available time series. This means that 
F2 now tests an \emph{interpolation} of the model rather than an 
\emph{extrapolation}. An advantage of doing this is that parameters such as the 
orbital period of the planet are best constrained by maximizing the baseline of 
available data and so this new approach allows for improved parameter estimates.

\section{RESULTS}
\label{sec:results}


\subsection{Overview}
\label{sub:overview}

In this work, all eight KOIs can be placed into one of three categories:
i) null detections (\S\ref{sub:nulls}), ii) spurious photometric detections 
(\S\ref{sub:photos}) and iii) spurious dynamical detections (\S\ref{sub:dynos}).
Each category is defined in the relevant subsection but it should be clear from
their naming that we find no confirmed or candidate exomoons in this sample.
For each KOI system studied, the TTVs and TDVs of all planetary candidates
can be seen in Figures~\ref{fig:TTV0463}-\ref{fig:TTV0784}. Parameter estimates 
from the marginalized posteriors of the favored light curve model are provided 
for all KOIs in Table~\ref{tab:finalparams}, except KOI-314 and KOI-784 which 
are available later in Table~\ref{tab:dynoparams}.

\begin{table*}
\caption{\emph{Final parameter estimates from the favored light curve model for 
KOIs studied in our sample, except KOI-314 and KOI-784 which are provided in
Table~\ref{tab:dynoparams}. In all cases, the favored model is a simple
planet-only model.
}} 
\centering 
\begin{tabular}{c c c c c c c} 
\hline
KOI & 463.01 & 3284.01 & 663.02 & 1596.02 & 494.01 & 252.01 \\ [0.5ex] 
\hline
$P$ \dotfill & $18.477637_{-0.000014}^{+0.000014}$ & $35.23266_{-0.00026}^{+0.00027}$ & $20.306531_{-0.000022}^{+0.000023}$ & $105.35794_{-0.00064}^{+0.00060}$ & $25.695907_{-0.000070}^{+0.000070}$ & $18.604627_{-0.000022}^{+0.00022}$ \\
$\tau$ [BKJD$_{\mathrm{UTC}}$] \dotfill & $758.07461_{-0.00033}^{+0.00033}$ & $769.5611_{-0.0034}^{+0.0036}$ & $781.85484_{-0.00049}^{+0.00049}$ & $665.4733_{-0.0023}^{+0.0023}$ & $779.7936_{-0.0012}^{+0.0012}$ & $874.68224_{-0.00052}^{+0.00055}$ \\
$(R_P/R_{\star})$ \dotfill & $0.0475_{-0.0010}^{+0.0011}$ & $0.01837_{-0.010}^{+0.011}$ & $0.02431_{-0.00055}^{+0.00105}$ & $0.0355_{-0.0018}^{+0.0020}$ & $0.03246_{-0.0095}^{+0.0107}$ & $0.0457_{-0.0022}^{+0.0035}$ \\ 
$\rho_{\star,\obs}$\,[g\,cm$^{-3}$] \dotfill & $27.0_{-8.5}^{+3.8}$ & $3.41_{-1.45}^{+0.98}$ & $7.2_{-2.4}^{+1.1}$ & $15.5_{-6.6}^{+3.7}$ & $3.64_{-1.35}^{+0.60}$ & $2.1_{-1.2}^{+1.5}$ \\
$b$ \dotfill & $0.29_{-0.20}^{+0.25}$ & $0.30_{-0.21}^{+0.32}$ & $0.31_{-0.21}^{+0.26}$ & $0.31_{-0.21}^{+0.31}$ & $0.30_{-0.21}^{+0.29}$ & $0.58_{-0.38}^{+0.21}$ \\
$q_1$ \dotfill & $0.22_{-0.13}^{+0.28}$ & $0.43_{-0.28}^{+0.36}$ & $0.42_{-0.14}^{+0.25}$ & $0.50_{-0.28}^{+0.32}$ & $0.27_{-0.17}^{+0.33}$ & $0.47_{-0.20}^{+0.32}$ \\
$q_2$ \dotfill & $0.29_{-0.21}^{+0.38}$ & $0.44_{-0.30}^{+0.35}$ & $0.62_{-0.26}^{+0.25}$ & $0.51_{-0.31}^{+0.31}$ & $0.37_{-0.25}^{+0.38}$ & $0.27_{-0.17}^{+0.36}$ \\
\hline
$R_P$\,[$R_{\oplus}$] \dotfill & $1.61_{-0.21}^{+0.21}$ & $1.04_{-0.10}^{+0.11}$ & $1.28_{-0.16}^{+0.17}$ & $1.90_{-0.29}^{+0.30}$ & $1.70_{-0.36}^{+0.36}$ & $2.56_{-0.20}^{+0.23}$ \\
$u_1$ \dotfill & $0.48_{-0.20}^{+0.18}$ & $0.78_{-0.40}^{+0.45}$ & $0.99_{-0.14}^{+0.13}$ & $0.91_{-0.38}^{+0.38}$ & $0.59_{-0.26}^{+0.25}$ & $0.74_{-0.25}^{+0.17}$ \\
$u_2$ \dotfill & $-0.03_{-0.18}^{+0.30}$ & $-0.18_{-0.30}^{+0.32}$ & $-0.36_{-0.11}^{+0.20}$ & $-0.26_{-0.26}^{+0.33}$ & $-0.10_{-0.19}^{+0.30}$ & $-0.03_{-0.29}^{+0.31}$ \\
$S_{\mathrm{eff}}$\,[$S_{\oplus}$] \dotfill & $0.896_{-0.098}^{+0.248}$ & $2.23_{-0.38}^{+0.99}$ & $3.12_{-0.35}^{+0.96}$ & $0.221_{-0.041}^{+0.100}$ & $3.54_{-0.80}^{+1.41}$ & $7.7_{-2.3}^{+5.4}$ \\
$(\rho_{\star,\obs}/\rho_{\star,\tru})$ \dotfill & $1.65_{-0.70}^{+0.97}$ & $0.59_{-0.26}^{+0.27}$ & $1.01_{-0.42}^{+0.57}$ & $2.3_{-1.2}^{+1.7}$ & $0.51_{-0.29}^{+0.51}$ & $0.36_{-0.19}^{+0.26}$ \\
$e_{\mathrm{min}}$ \dotfill & $0.18_{-0.12}^{+0.14}$ & $0.18_{-0.13}^{+0.18}$ & $0.00_{-0.00}^{+0.17}$ & $0.00_{-0.00}^{+0.17}$ & $0.22_{-0.22}^{+0.25}$ & $0.33_{-0.17}^{+0.21}$ \\ 
$(M_S/M_P)$ \dotfill  & $<0.92$ & $<0.29$ & $<0.52$ & $<0.60$ & $<0.66$ & $<0.32$ \\ [1ex]
\hline\hline 
\end{tabular}
\label{tab:finalparams} 
\end{table*}

\subsection{Null Detections}
\label{sub:nulls}

The first category of objects we discuss in this section are the null 
detections. We define these objects to be ones for which the mode of the derived 
satellite-to-planet mass ratio $(M_S/M_P)$ posterior distribution is at, or 
close to, zero plus other detection criteria are also failed. Note that this is 
distinct from requiring that the objects fail criterion B4a, although 
any object in this category will indeed fail B4a too. We find five objects 
in this category: KOI-663.02, KOI-1596.02, KOI-494.01, KOI-463.01 and 
KOI-3284.01. Null detections, such as these, allow for a simple calculation of 
the upper limit on the ratio $(M_S/M_P)$ by posterior marginalization. These 
upper limits range from 0.29 to 0.92 (95\% confidence upper limits) and are
available in Table~\ref{tab:finalparams}.

\begin{table*}
\caption{\emph{Detection criteria results for each of the eight KOIs surveyed
in this work. Only KOI-314.02 and KOI-784.01 pass the five criteria listed
here.
}} 
\centering 
\begin{tabular}{l c c c c c c c l} 
\hline
KOI & B1 & B2a & B3a & B4a & \,\,\,\,\,\,\,\,\,F2 \\ [0.5ex] 
\hline
KOI-463.01 & \text{\sffamily X}\,($+2.89$\,$\sigma$) & \text{\sffamily X}\,($-0.33$) & \checkmark\,(4.82\%) & \text{\sffamily X}\,(FAP=29.86\%) & \text{\sffamily X}\,($\Delta\chi^2=27.8$ for $N=1815$) \\ 
KOI-314.02 & \checkmark\,($+14.3$\,$\sigma$) & \checkmark\,($+0.28$) & \checkmark\,(0.00\%) & \checkmark\,(FAP=0.066\%) & \checkmark\,($\Delta\chi^2=-925.0$ for $N=9161$) \\ 
KOI-784.01 & \checkmark\,($+6.50$\,$\sigma$) & \checkmark\,($+0.32$) & \checkmark\,(0.00\%) & \checkmark\,(FAP=0.018\%) & \checkmark\,($\Delta\chi^2=-39.0$ for $N=9431$) \\ 
KOI-3284.01 & \text{\sffamily X}\,($+3.32$\,$\sigma$) & \checkmark\,($+0.58$) & \text{\sffamily X}\,($94.2$\%) & \text{\sffamily X}\,(FAP=37.28\%) & \text{\sffamily X}\,($\Delta\chi^2=15.8$ for $N=597$) \\ 
KOI-663.02 & \text{\sffamily X}\,($+0.76$\,$\sigma$) & \text{\sffamily X}\,($-0.44$) & \text{\sffamily X}\,($38.8$\%) & \text{\sffamily X}\,(FAP=47.61\%) & \text{\sffamily X}\,($\Delta\chi^2=12.8$ for $N=9676$) \\ 
KOI-1596.02 & \text{\sffamily X}\,($+1.62$\,$\sigma$) & \checkmark\,($+0.34$) & \checkmark\,(2.09\%) & \text{\sffamily X}\,(FAP=34.13\%) & \checkmark\,($\Delta\chi^2=-4.1$ for $N=322$) \\ 
KOI-494.01 & \text{\sffamily X}\,($+1.23$\,$\sigma$) & \text{\sffamily X}\,($-0.46$) & \text{\sffamily X}\,($39.6$\%) & \text{\sffamily X}\,(FAP=30.55\%) & \text{\sffamily X}\,($\Delta\chi^2=16.0$ for $N=836$)\\ 
KOI-252.01 & \text{\sffamily X}\,($+1.82$\,$\sigma$) & \checkmark\,($+0.52$) & \checkmark\,(0.40\%) & \text{\sffamily X}\,(FAP=16.13\%) & \checkmark\,($\Delta\chi^2=-4.9$ for $N=33048$) \\ [1ex] 
\hline\hline 
\end{tabular}
\label{tab:criteria} 
\end{table*}

As well as failing criteria B4a, all of these objects fail at least one other
detection criteria, as listed in Table~\ref{tab:criteria}. For several cases,
a periodogram search of the TTVs and TDVs for these objects appears to indicate 
possible perturbations, as shown in Figures~\ref{fig:TTV0463}-\ref{fig:TTV0494}. 
We have listed the most significant peaks from a periodogram search in 
Table~\ref{tab:timing}.

\ifthenelse{\boolean{color}}{
\begin{figure*}
\begin{center}
\includegraphics[width=16.8 cm]{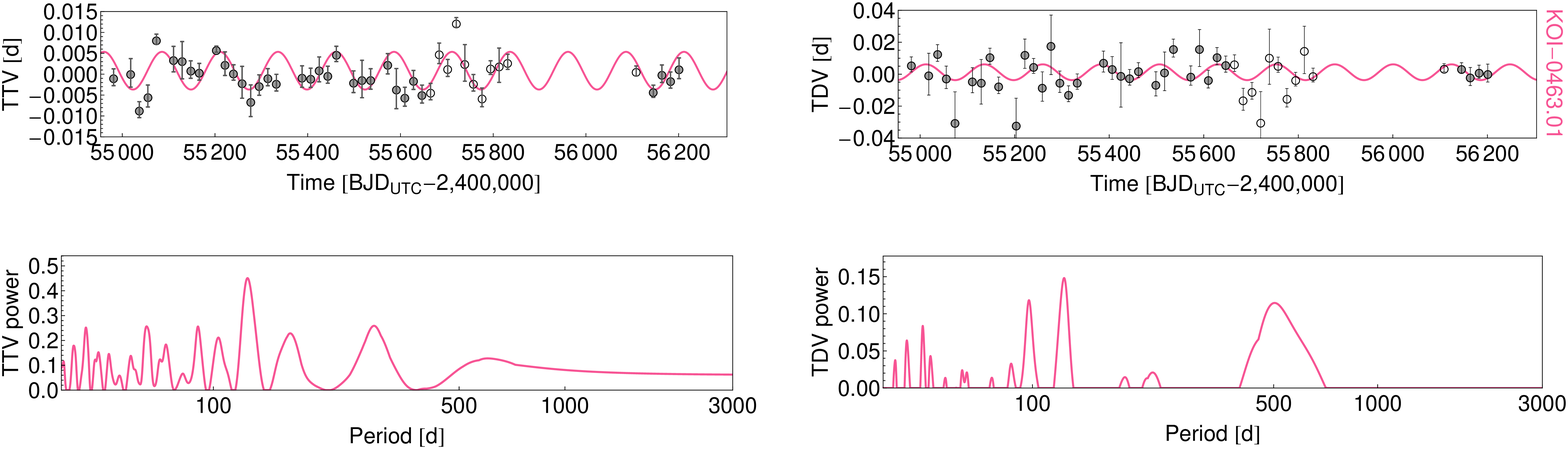}
\caption{\emph{TTVs (left) and TDVs (right) for the planetary candidate of
KOI-463. Points used in the photodynamical moon fits are filled circles, 
whereas those points ignored for subsequent predictive tests are open circles.
Bottom panels show the periodograms of the TTVs \& TDVs (which use the entire
available data set).}} 
\label{fig:TTV0463}
\end{center}
\end{figure*}
}{
\begin{figure*}
\begin{center}
\includegraphics[width=16.8 cm]{final.0463.bw.eps}
\caption{\emph{TTVs (left) and TDVs (right) for the planetary candidate of
KOI-463. Points used in the photodynamical moon fits are filled circles, 
whereas those points ignored for subsequent predictive tests are open circles.
Bottom panels show the periodograms of the TTVs \& TDVs (which use the entire
available data set).}} 
\label{fig:TTV0463}
\end{center}
\end{figure*}
}

\ifthenelse{\boolean{color}}{
\begin{figure*}
\begin{center}
\includegraphics[width=16.8 cm]{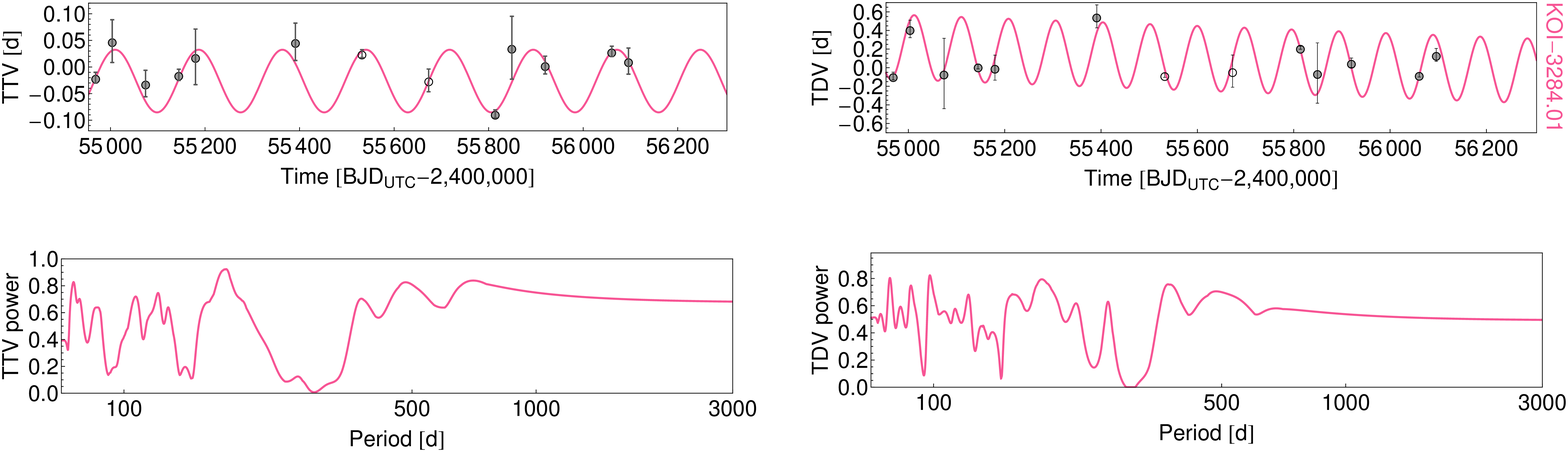}
\caption{\emph{TTVs (left) and TDVs (right) for the planetary candidate of
KOI-3284. Points used in the photodynamical moon fits are filled circles, 
whereas those points ignored for subsequent predictive tests are open circles.
Bottom panels show the periodograms of the TTVs \& TDVs (which use the entire
available data set).}} 
\label{fig:TTV3284}
\end{center}
\end{figure*}
}{
\begin{figure*}
\begin{center}
\includegraphics[width=16.8 cm]{final.3284.bw.eps}
\caption{\emph{TTVs (left) and TDVs (right) for the planetary candidate of
KOI-3284. Points used in the photodynamical moon fits are filled circles, 
whereas those points ignored for subsequent predictive tests are open circles.
Bottom panels show the periodograms of the TTVs \& TDVs (which use the entire
available data set).}} 
\label{fig:TTV3284}
\end{center}
\end{figure*}
}

\ifthenelse{\boolean{color}}{
\begin{figure*}
\begin{center}
\includegraphics[width=16.8 cm]{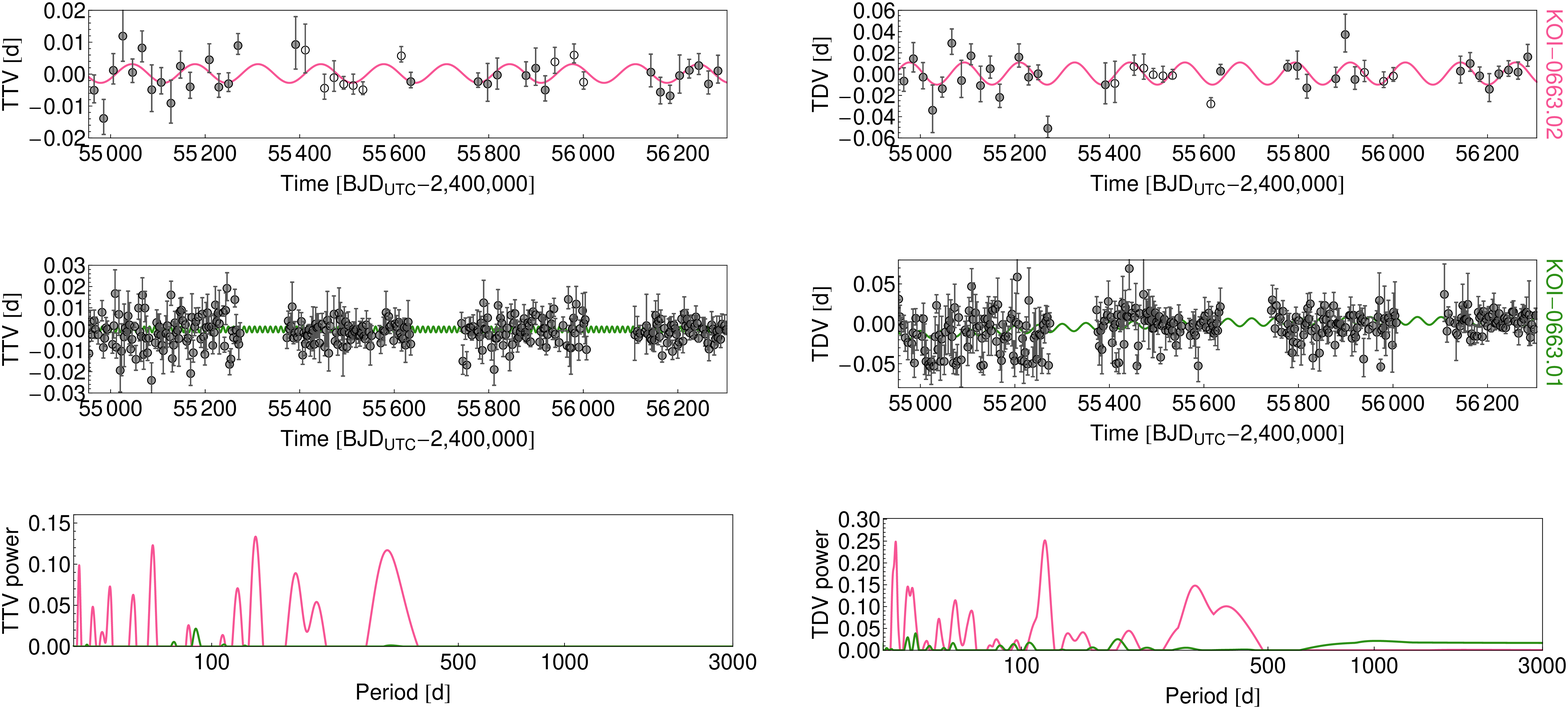}
\caption{\emph{TTVs (left) and TDVs (right) for the planetary candidates of
KOI-663. Each row shows the observations (circles) with the best-fit 
sinusoid from a periodogram using the same coloring as that for which the name
of the planet is highlighted in on the RHS. For KOI-663.02, points used in the 
photodynamical moon fits are filled circles, whereas those points ignored for 
subsequent predictive tests are open circles. Bottom panels show the 
periodograms of the TTVs \& TDVs (which use the entire available data set), with 
vertical grid lines marking the locations where one might expect power from 
planet-planet interactions.}} 
\label{fig:TTV0663}
\end{center}
\end{figure*}
}{
\begin{figure*}
\begin{center}
\includegraphics[width=16.8 cm]{final.0663.bw.eps}
\caption{\emph{TTVs (left) and TDVs (right) for the planetary candidates of
KOI-663. Each row shows the observations (circles) with the best-fit 
sinusoid from a periodogram using consistent shading for each planet. For 
KOI-663.02, points used in the photodynamical moon fits are filled circles, 
whereas those points ignored for subsequent predictive tests are open circles. 
Bottom panels show the periodograms of the TTVs \& TDVs (which use the entire 
available data set), with vertical grid lines marking the locations where one 
might expect power from planet-planet interactions.}} 
\label{fig:TTV0663}
\end{center}
\end{figure*}
}

\ifthenelse{\boolean{color}}{
\begin{figure*}
\begin{center}
\includegraphics[width=16.8 cm]{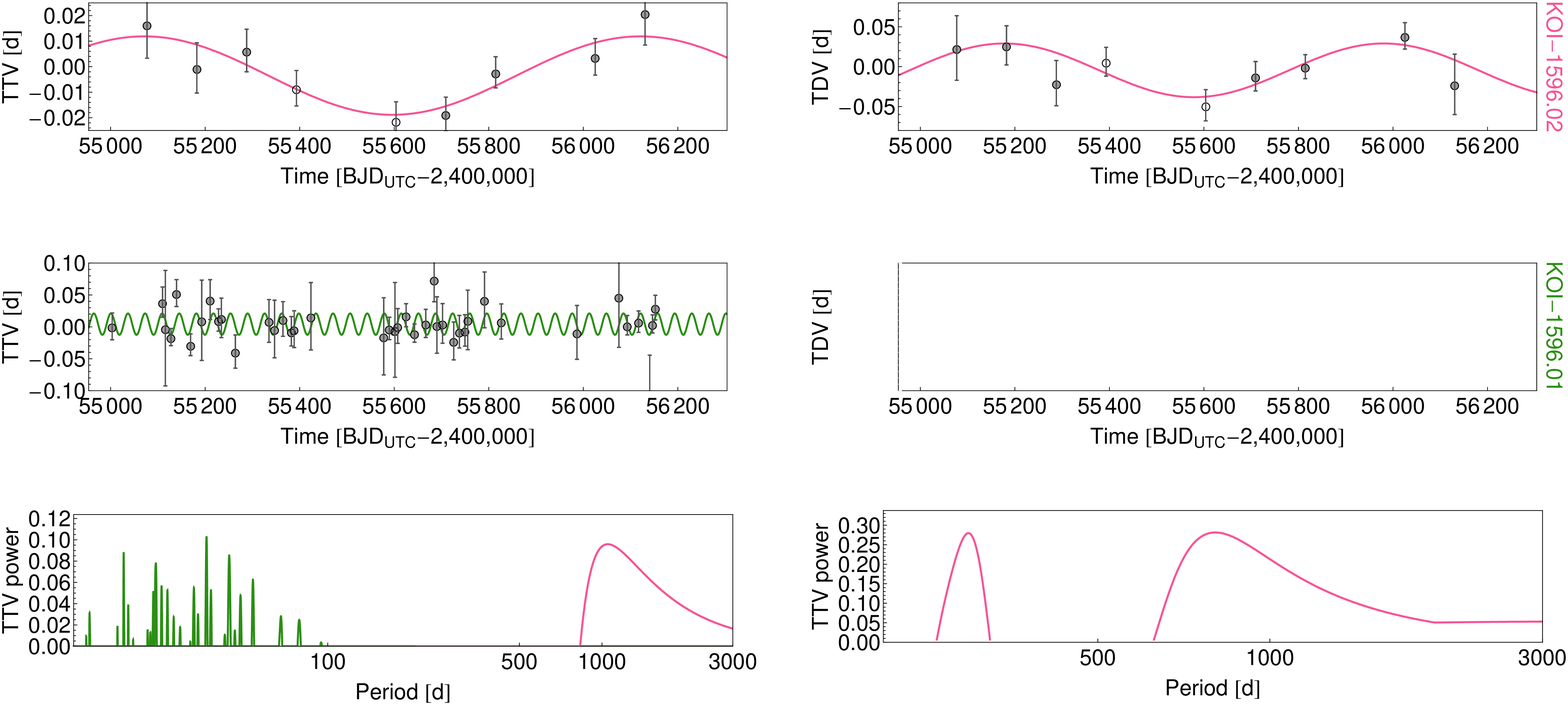}
\caption{\emph{TTVs (left) and TDVs (right) for the planetary candidates of
KOI-1596. Each row shows the observations (circles) with the best-fit 
sinusoid from a periodogram using the same coloring as that for which the name
of the planet is highlighted in on the RHS. For KOI-1596.02, points used in the 
photodynamical moon fits are filled circles, whereas those points ignored for 
subsequent predictive tests are open circles. Bottom panels show the 
periodograms of the TTVs \& TDVs (which use the entire available data set), with 
vertical grid lines marking the locations where one might expect power from 
planet-planet interactions.}} 
\label{fig:TTV1596}
\end{center}
\end{figure*}
}{
\begin{figure*}
\begin{center}
\includegraphics[width=16.8 cm]{final.1596.bw.eps}
\caption{\emph{TTVs (left) and TDVs (right) for the planetary candidates of
KOI-1596. Each row shows the observations (circles) with the best-fit 
sinusoid from a periodogram using consistent shading for each planet. For 
KOI-1596.02, points used in the photodynamical moon fits are filled circles, 
whereas those points ignored for subsequent predictive tests are open circles. 
Bottom panels show the periodograms of the TTVs \& TDVs (which use the entire 
available data set), with vertical grid lines marking the locations where one 
might expect power from planet-planet interactions.}} 
\label{fig:TTV1596}
\end{center}
\end{figure*}
}

\ifthenelse{\boolean{color}}{
\begin{figure*}
\begin{center}
\includegraphics[width=16.8 cm]{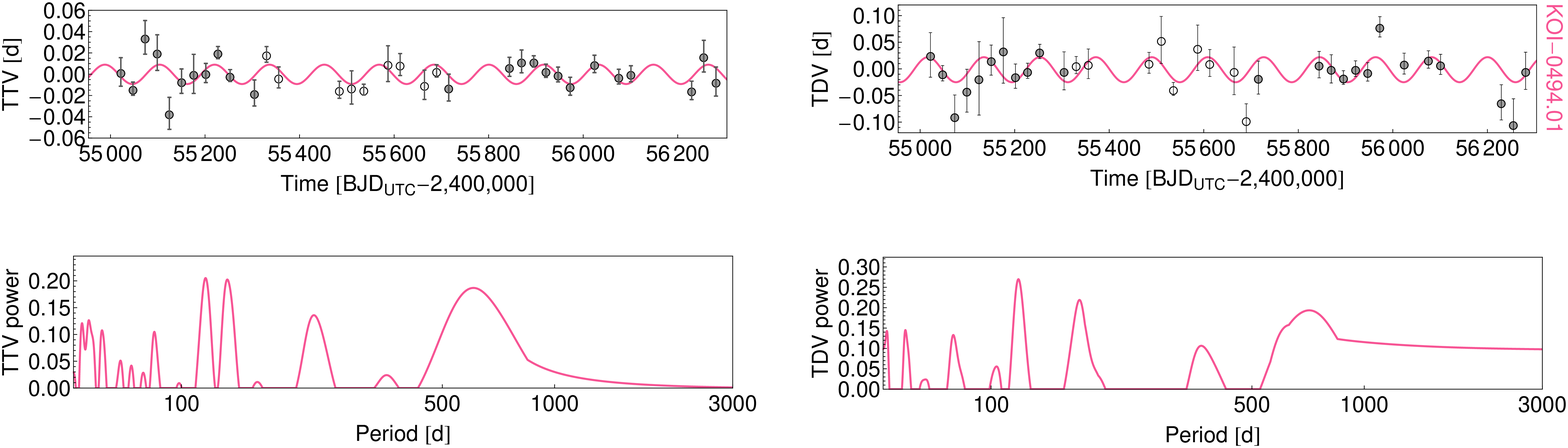}
\caption{\emph{TTVs (left) and TDVs (right) for the planetary candidate of
KOI-494. Points used in the photodynamical moon fits are filled circles, 
whereas those points ignored for subsequent predictive tests are open circles.
Bottom panels show the periodograms of the TTVs \& TDVs (which use the entire
available data set).}} 
\label{fig:TTV0494}
\end{center}
\end{figure*}
}{
\begin{figure*}
\begin{center}
\includegraphics[width=16.8 cm]{final.0494.bw.eps}
\caption{\emph{TTVs (left) and TDVs (right) for the planetary candidate of
KOI-494. Points used in the photodynamical moon fits are filled circles, 
whereas those points ignored for subsequent predictive tests are open circles.
Bottom panels show the periodograms of the TTVs \& TDVs (which use the entire
available data set).}} 
\label{fig:TTV0494}
\end{center}
\end{figure*}
}

KOI-463.01 and KOI-3284.01 are two single KOI systems which show $>4$\,$\sigma$ 
significant TTVs (from an F-test). This may therefore be evidence for additional
planets although we consider the signal-to-noise of the available data 
insufficient to reliably deduce a unique unseen perturber solution. We consider 
the TTVs/TDVs of the only other single KOI classed as an exomoon null-detection, 
KOI-494.01, to be insignificant.

Two out of the five null detection cases reside in multi-transiting planet
systems where interactions may be a-priori expected. However, the period
ratios between KOI-663.02/KOI-663.01 and KOI-1596.02/KOI-1596.01 are 7.4 and
17.8 respectively i.e. too far to expect significant interactions. This point
is reinforced empirically by noting that the periodograms show no strong
overlapping peaks and no significant ($\geq4$\,$\sigma$) power. We are therefore
unable to confirm these two multi-planetary systems using TTVs.

\begin{table*}
\caption{\emph{Summary of the highest power peaks found in the periodograms
of the TTVs and TDVs of the planetary candidates analyzed in this work.
}} 
\centering 
\begin{tabular}{c c c c c c c} 
\hline
KOI & TTV Period [d] & TTV Amp. [mins] & TTV Signif. & TDV Period [d] & TDV Amp. [mins] & TDV Signif. \\ [0.5ex] 
\hline
KOI-463.01 & 125.1 & 6.5 & 4.8\,$\sigma$ & 123.6 & 7.1 & 2.3\,$\sigma$ \\ 
KOI-314.02 & 112.9 & 21.7 & 9.3\,$\sigma$ & 307.3 & 6.9 & 3.9\,$\sigma$ \\ 
KOI-314.01 & 913.8 & 2.5 & 3.8\,$\sigma$ & 38.7 & 3.1 & 2.4\,$\sigma$ \\ 
KOI-314.03 & 434.0 & 16.4 & 3.2\,$\sigma$ & - & - & - \\ 
KOI-784.01 & 541.9 & 25.6 & 4.9\,$\sigma$ & 53.9 & 54.1 & 4.8\,$\sigma$ \\ 
KOI-784.02 & 29.0 & 7.2 & 3.3\,$\sigma$ & - & - & - \\ 
KOI-3284.01 & 176.9 & 84.8 & 5.3\,$\sigma$ & 97.9 & 5.0 & 3.3\,$\sigma$ \\ 
KOI-663.02 & 133.1 & 4.2 & 2.5\,$\sigma$ & 116.6 & 15.0 & 2.9\,$\sigma$ \\ 
KOI-663.01 & 11.1 & 2.0 & 2.9\,$\sigma$ & 50.1 & 7.1 & 5.4\,$\sigma$ \\ 
KOI-1596.02 & 1050.0 & 22.1 & 2.7\,$\sigma$ & 802.7 & 48.5 & 1.7\,$\sigma$ \\ 
KOI-1596.01 & 36.1 & 24.1 & 2.7\,$\sigma$ & - & - & - \\ 
KOI-494.01 & 116.1 & 13.1 & 2.6\,$\sigma$ & 116.1 & 33.9 & 2.7\,$\sigma$ \\ 
KOI-252.01 & 128.4 & 3.3 & 3.4\,$\sigma$ & 81.1 & 12.9 & 2.9\,$\sigma$ \\ [1ex] 
\hline\hline 
\end{tabular}
\label{tab:timing} 
\end{table*}

\subsection{Spurious Photometric Detections}
\label{sub:photos}

A spurious photometric detection is one where we hypothesize that a residual
amount of time-correlated noise drives a non-zero satellite radius solution,
which then itself drives a non-zero satellite mass solution due to the fact
we enforce only physically reasonable satellite densities in the fits. We
identify such cases by the fact that: i) the posterior distribution for
$(M_S/M_P)$ does not peak at zero (and thus it is not a ``null'' detection), 
ii) other detection criteria are failed (suggesting some kind of spurious
detection) and iii) performing a fit where the moon's radius is fixed to zero 
(model $\mathcal{S}_{R0}$) yields an $(M_S/M_P)$ posterior which peaks at zero, 
as expected for a null detection. This final point indicates that when we switch 
off the moon's radius, the planet no longer seems to be perturbed and thus the 
moon's radius was likely a result of the regression fitting out 
non-astrophysical artifacts in the light curve. KOI-252.01 is the only object 
in this survey which we classify in this category.

\ifthenelse{\boolean{color}}{
\begin{figure*}
\begin{center}
\includegraphics[width=16.8 cm]{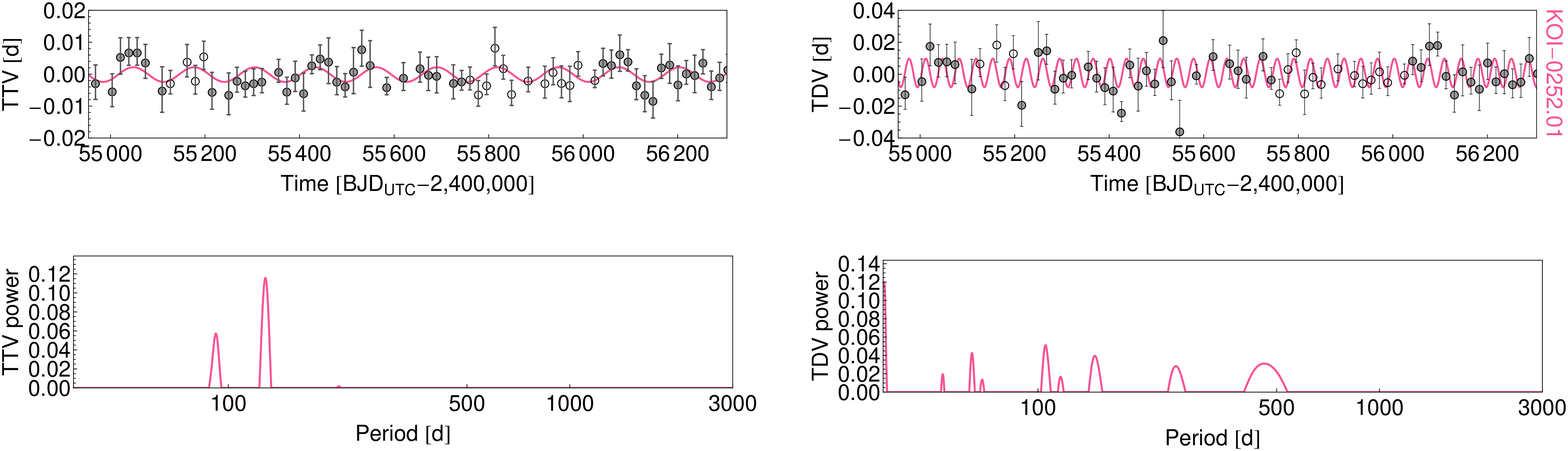}
\caption{\emph{TTVs (left) and TDVs (right) for the planetary candidate of
KOI-252. Points used in the photodynamical moon fits are filled circles, 
whereas those points ignored for subsequent predictive tests are open circles.
Bottom panels show the periodograms of the TTVs \& TDVs (which use the entire
available data set).}} 
\label{fig:TTV0252}
\end{center}
\end{figure*}
}{
\begin{figure*}
\begin{center}
\includegraphics[width=16.8 cm]{final.0252.bw.eps}
\caption{\emph{TTVs (left) and TDVs (right) for the planetary candidate of
KOI-252. Points used in the photodynamical moon fits are filled circles, 
whereas those points ignored for subsequent predictive tests are open circles.
Bottom panels show the periodograms of the TTVs \& TDVs (which use the entire
available data set).}} 
\label{fig:TTV0252}
\end{center}
\end{figure*}
}

As seen in Table~\ref{tab:criteria}, B1 and B4a are both failed for KOI-252.01.
The latter is because the derived $(M_S/M_P)$ peaks away from zero (at around
0.5) and yet is extremely disperse over the entire prior. The former essentially
implies that whatever is driving the fit is actually of relatively low 
significance.

By definition of being a spurious photometric detection, the $(M_S/M_P)$ 
posterior has a mode at zero when we enforce $(R_S/R_P)=0$ in the 13-dimensional
model $\mathcal{S}_{R0}$. The results from this model are now consistent with a 
null detection and thus we may derive upper limits on $(M_S/M_P)$ as usual (this
is similar to the procedure used for KOI-303.01 in \citealt{hek:2013}). For 
KOI-252.01 then, we derive that $(M_S/M_P)<0.33$ to 95\% confidence.

\subsection{Spurious Dynamical Detections}
\label{sub:dynos}

KOI-314.02 and KOI-784.01 are the only two remaining KOIs and we classify
these both as spurious dynamical detections. A-priori, these two systems have
the highest chance of exhibiting significant planet-planet interactions since
KOI-314.02 and KOI-314.01 reside near a 5:3 period commensurability and 
KOI-784.01 and KOI-784.02 are close to 2:1. In the case of KOI-314, there is a
third planetary candidate interior to the other two, for which the closest
commensurabilities are 9:4 and 5:4 relative to KOI-314.02 and KOI-314.01
respectively. Given that the 9:4 commensurabily does not usually exhibit strong 
perturbations and the fact KOI-314.03 is three times smaller than the other
two KOIs, the lack of any observed interactions induced by this candidate is
not surprising.

\ifthenelse{\boolean{color}}{
\begin{figure*}
\begin{center}
\includegraphics[width=16.8 cm]{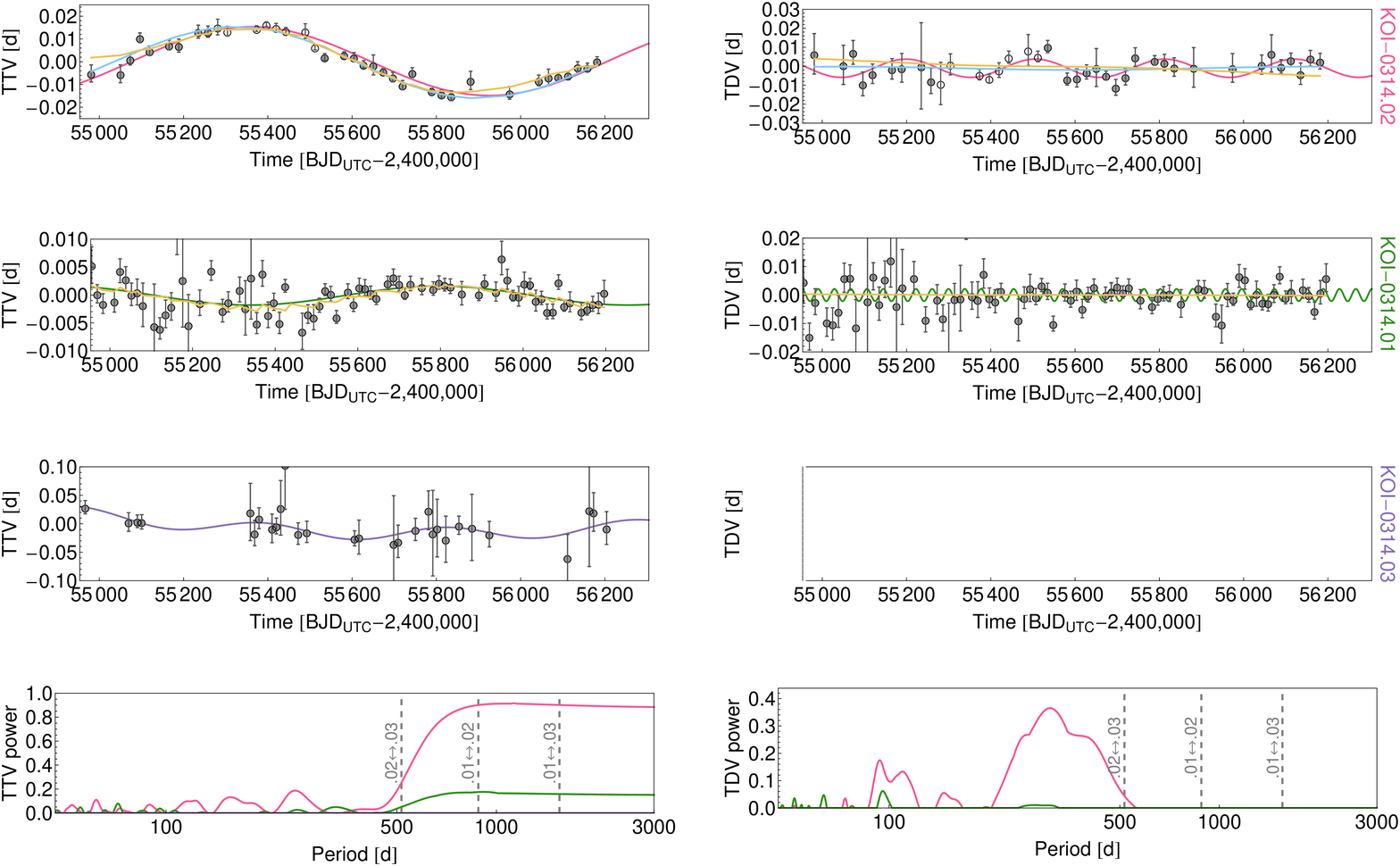}
\caption{\emph{TTVs (left) and TDVs (right) for the planetary candidates of
KOI-314. Each row shows the observations (circles) with the best-fit 
sinusoid from a periodogram using the same coloring as that for which the name
of the planet is highlighted in on the RHS. Additionally, we show the best-fit
moon model in blue and best-fit planet-planet model in orange.
For KOI-314.02, points used in the photodynamical moon fits are filled 
circles, whereas those points ignored for subsequent predictive tests are open 
circles. Bottom panels show the periodograms of the TTVs \& TDVs (which use the 
entire available data set), with vertical grid lines marking the locations where
one might expect power from planet-planet interactions.}} 
\label{fig:TTV0314}
\end{center}
\end{figure*}
}{
\begin{figure*}
\begin{center}
\includegraphics[width=16.8 cm]{final.0314.bw.eps}
\caption{\emph{TTVs (left) and TDVs (right) for the planetary candidates of
KOI-314. Each row shows the observations (circles) with the best-fit 
sinusoid from a periodogram using the consistent shading for each planet. 
Additionally, we show the best-fit moon model in gray-solid and best-fit 
planet-planet model in gray-dashed. For KOI-314.02, points used in the 
photodynamical moon fits are filled circles, whereas those points ignored for 
subsequent predictive tests are open circles. Bottom panels show the 
periodograms of the TTVs \& TDVs (which use the entire available data set), with 
vertical grid lines marking the locations where one might expect power from 
planet-planet interactions.}} 
\label{fig:TTV0314}
\end{center}
\end{figure*}
}

\ifthenelse{\boolean{color}}{
\begin{figure*}
\begin{center}
\includegraphics[width=16.8 cm]{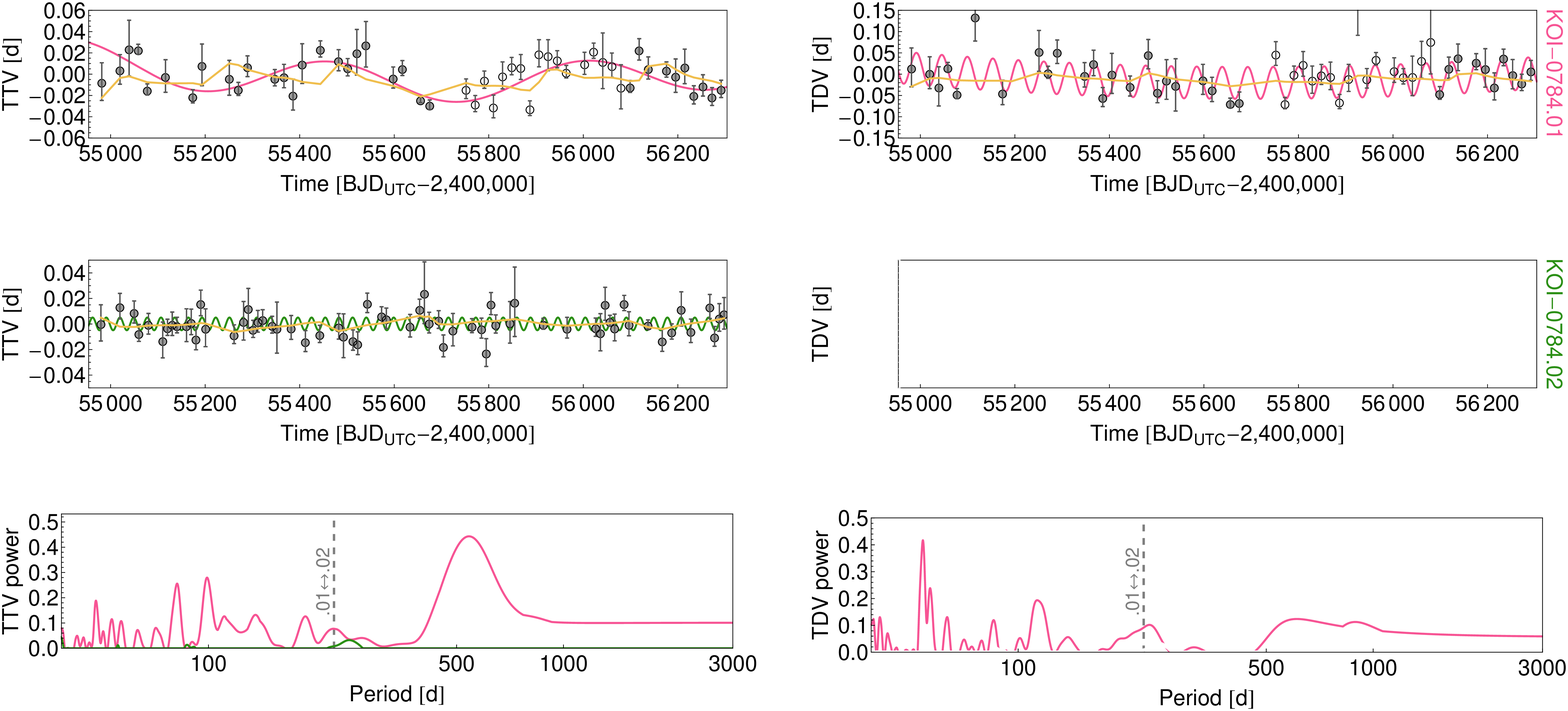}
\caption{\emph{TTVs (left) and TDVs (right) for the planetary candidates of
KOI-784. Each row shows the observations (circles) with the best-fit 
sinusoid from a periodogram using the same coloring as that for which the name
of the planet is highlighted in on the RHS. Additionally, we show the best-fit
planet-planet model in orange. For KOI-784.01, points used in the 
photodynamical moon fits are filled circles, whereas those points ignored for 
subsequent predictive tests are open circles. Bottom panels show the 
periodograms of the TTVs \& TDVs (which use the entire available data set), with 
vertical grid lines marking the locations where one might expect power from 
planet-planet interactions.}} 
\label{fig:TTV0784}
\end{center}
\end{figure*}
}{
\begin{figure*}
\begin{center}
\includegraphics[width=16.8 cm]{final.0784.bw.eps}
\caption{\emph{TTVs (left) and TDVs (right) for the planetary candidates of
KOI-784. Each row shows the observations (circles) with the best-fit 
sinusoid from a periodogram using consistent shading for each planet. 
Additionally, we show the best-fit planet-planet model in orange. For 
KOI-784.01, points used in the photodynamical moon fits are filled circles, 
whereas those points ignored for subsequent predictive tests are open circles. 
Bottom panels show the periodograms of the TTVs \& TDVs (which use the entire 
available data set), with vertical grid lines marking the locations where one 
might expect power from planet-planet interactions.}} 
\label{fig:TTV0784}
\end{center}
\end{figure*}
}

\begin{figure}
\begin{center}
\includegraphics[width=8.4 cm]{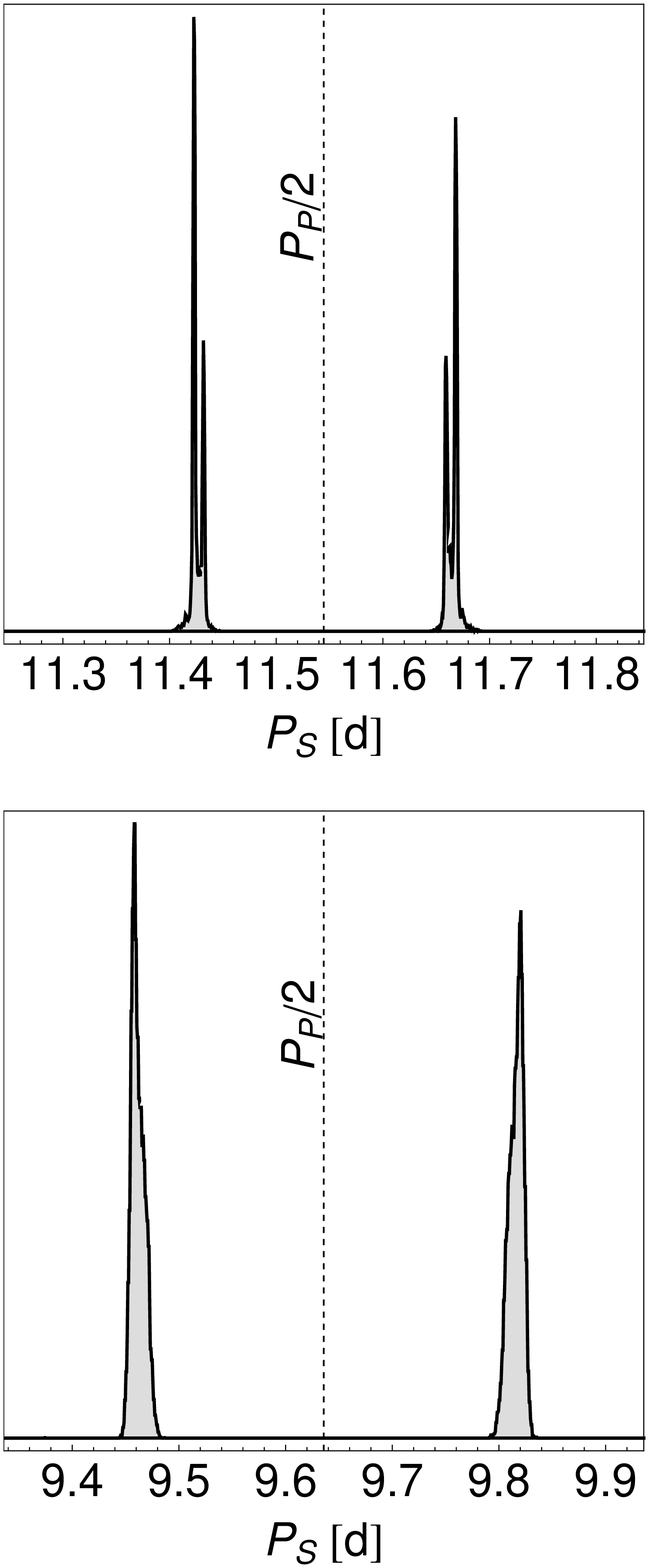}
\caption{\emph{Posterior distribution for a putative moon's orbital period
for KOI-314.02 (upper panel) and KOI-784.01 (lower panel). In both cases,
the observed long-period TTV requires a finely tuned moon period close to
a 2:1 commensurability with the planet, which we deem quite improbable.}} 
\label{fig:Pshistos}
\end{center}
\end{figure}

For both KOI-314.02 and KOI-784.01, the moon model $\mathcal{S}$ yields
an $(M_S/M_P)$ posterior peaking at unity (i.e. the solutions favor binary
planets). Formally, all of the detection criteria are satisfied despite this, as 
seen in Table~\ref{tab:criteria}. From the TTVs shown in 
Figures~\ref{fig:TTV0314} \& \ref{fig:TTV0784} and the list of periodogram 
frequencies detailed in Table~\ref{tab:timing}, it is clear that significant 
power exists at long-periods for both objects. Since exomoons always have an 
orbital period less than that of the planet they are bound to 
\citep{kipping:2009a}, the moon interpretation requires these to be aliases of
the true short period. However, producing such a long period requires fine
tuning of the moon's period. This is easily visualized in 
Figure~\ref{fig:Pshistos}, where one can see that for both cases we require a
moon's period to be nearly 1:2 commensurable to the planet's period, with a 
small frequency splitting which induces the longer super-period. Such fine 
tuning is a strong blow to the moon hypothesis since there is no known example 
of a Solar System moon being in such a near-commensurability.

A crude but useful tool in studying TTVs of multi-planet systems is to estimate 
the ``super-period'' of the TTVs assuming the variations are sinusoidal
\citep{xie:2013}. Consider two planets with orbital periods $P'$ and $P$ where 
$P'>P$. For two planets in a $k^{\mathrm{th}}$ order MMR, the ratio of the 
orbital periods is simply $(P'/P) \simeq j/(j-k)$, where $k$ is an integer 
satisfying $0<k<j$ and $j$ defines the mutual proximity of the MMR. Following
the approximate theory of \citet{xie:2013}, the expected periodicity of the 
observed TTVs, the so-called ``super-period'', is given by:

\begin{align}
\Delta &\equiv \Bigg(\frac{P'}{P}\Bigg) \Bigg(\frac{j-k}{j}\Bigg) - 1,\\
P_{\mathrm{TTV}} &= \frac{P'}{j |\Delta|}.
\end{align}

For the KOI-314 and KOI-784 systems, we have marked $P_{\mathrm{TTV}}$ on
the periodograms shown in Figures~\ref{fig:TTV0314} \& \ref{fig:TTV0784}. The
pair KOI-314.01/.02 indeed shows power at this frequency and performing a full 
dynamical fit of the TTVs (described in the Appendix) retrieves a 
quasi-sinusoidal signal conforming with this frequency. In contrast, our 
dynamical fits of KOI-784.01/.02 reveal a non-sinusoidal waveform, which 
explains why the expected super-period has no accompanying power in the 
periodogram. 

In both systems, the TTVs fitted by our moon model $\mathcal{S}$ produce an
essentially equally good fit to the data as a planet-planet interaction model.
This is clearly evident by comparing the fits shown in Figures~\ref{fig:TTV0314} 
\& \ref{fig:TTV0784}. However, the moon model requires an entirely new object
to be introduced into the system to explain the observations, whereas the 
planet-planet models naturally explain the TTVs simply using the known 
transiting planets. This fact, combined with the previously discussed concerns 
with the moon hypothesis, leads us to conclude that both KOI-314 and KOI-784 
are spurious dynamical detections i.e. a planet-planet perturbation is the most 
likely underlying reason via Occam's Razor.

For both KOI-314 and KOI-784, we conducted two sets of dynamical TTV fits 
where we turned on/off the mass of the planets. Since our fits are conducted 
with \multi, we may compare the Bayesian evidences to evaluate the statistical 
significance of the claimed interactions, for which we find healthy confidences
of $24.2$\,$\sigma$ and $6.7$\,$\sigma$ for KOI-314.01/.02 and KOI-784.01/.02 
respectively. The maximum a-posteriori fit through the TTVs \& TDVs (where
available) reduces the $\chi^2$ from $1440.1\rightarrow335.2$ with 218 
observations for KOI-314.01/.02 and $487.7\rightarrow340.0$ with 166 
observations for KOI-784.01/.02. A simple F-test based on these $\chi^2$
changes finds confidences of $17.1$\,$\sigma$ and $7.1$\,$\sigma$ for
KOI-314.01/.02 and KOI-784.01/.02 respectively, showing good consistency with 
the Bayesian evidence calculation.

Statistics aside, Figure~\ref{fig:TTV0314} shows that our fits for 
KOI-314.01/02 clearly reproduce the dominant pattern in the data. The two sets 
of TTVs exhibit visible anti-correlation, which is a well-known signature of 
planet-planet interactions \citep{steffen:2012} and so further improves our 
confidence in this fit. Finally, we note that the derived solution is
dynamically stable for $\geq$\,Gyr and yields physically plausible internal
compositions for both objects (see Table~\ref{tab:dynoparams}). 

Although \multi\ only identifies one mode and thus the solution appears unique, 
the unweighted posteriors show high likelihood solutions extending down a thin
tail of higher eccentricities and higher masses. These solutions have much lower
sample probability since they require the fine tuning of $\Delta\varpi\simeq0$ 
to explain the data. Given our assumption of uniform priors in $\varpi$, the 
likelihood multiplied by the prior mass, which defines the sample probability, 
is much lower along this tail and thus has a much lower Bayesian evidence. For
this reason, this tail does not appear when \multi\ outputs the weighted
posteriors. We decided to explore this tail in a second fit by enforcing a
lower limit of $(M_P/M_{\star})>1.5\times10^{-5}$. The mode which is picked up
by \multi\ down this tail is easily identified as unphysical, since it requires 
that 77.4\% of KOI-314.01's posterior samples exceed the mass-stripping limit of 
\citet{marcus:2009} (i.e. the density is unphysically high). The fact the
solution is both unphysical and has a much lower sample probability allows us to 
discard it.

We therefore consider the original solution to be the only plausible explanation 
for the TTVs. We subsequently consider these planetary candidates to be 
confirmed as exhibiting interactions and thus are real planets given the low 
derived masses. We refer to KOI-314.01 and .02 as KOI-314b and KOI-314c 
respectively from here on. Note that KOI-314.03 exhibits no detectable signature 
in any of the TTVs and thus this object remains a planetary candidate at this 
time.

\ifthenelse{\boolean{color}}{
\begin{figure*}
\begin{center}
\includegraphics[width=16.8 cm]{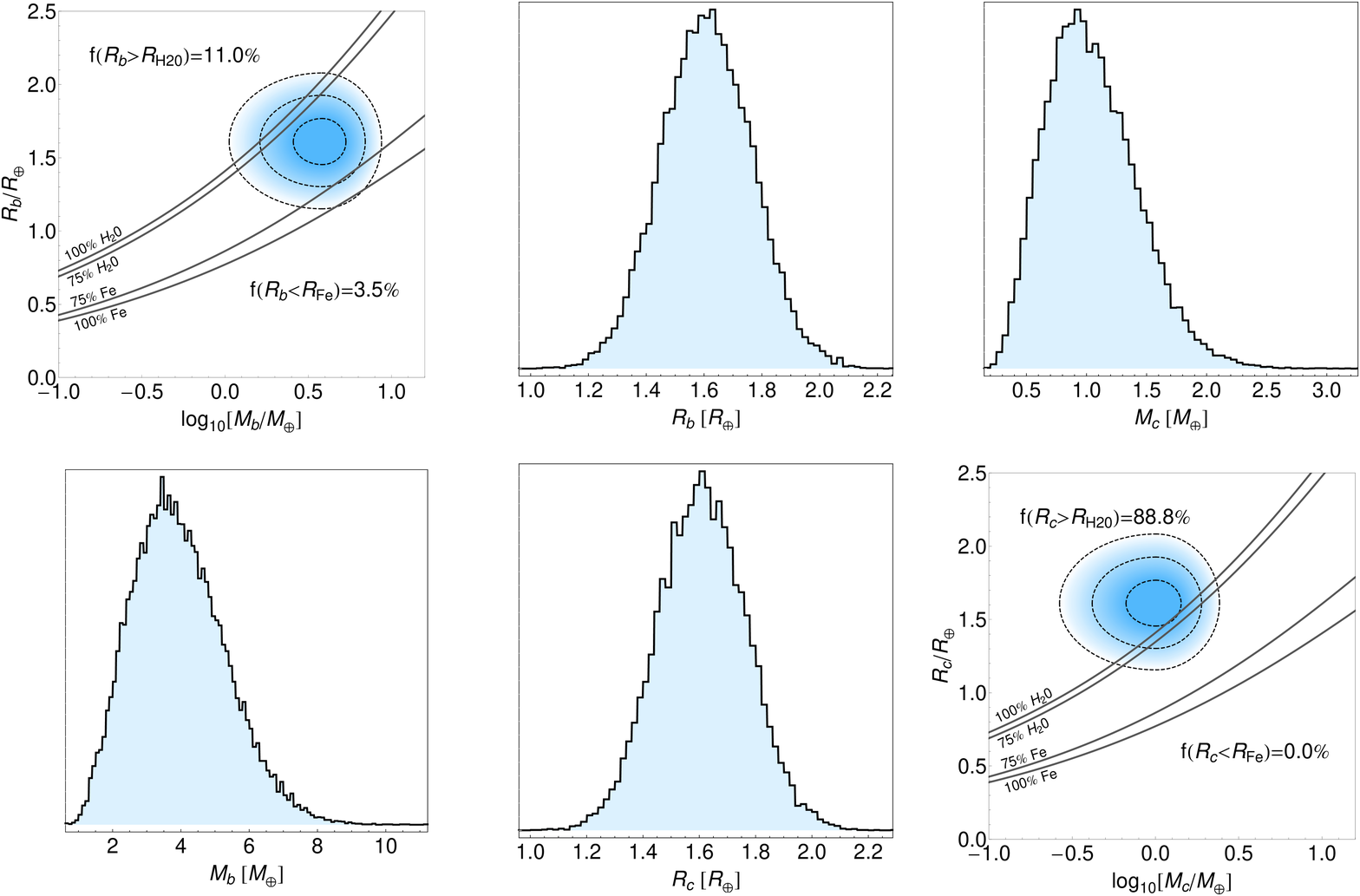}
\caption{\emph{Posterior distributions for the mass and radius of planets
KOI-314b and KOI-314c. The joint posteriors in the corners reveal that
the inner planet, b, is likely rocky, whereas the Earth-mass outer planet, c, 
likely maintains an extended atmosphere comprising $\geq17_{-13}^{+12}$\% of the 
total radius.}} 
\label{fig:GRID0314}
\end{center}
\end{figure*}
}{
\begin{figure*}
\begin{center}
\includegraphics[width=16.8 cm]{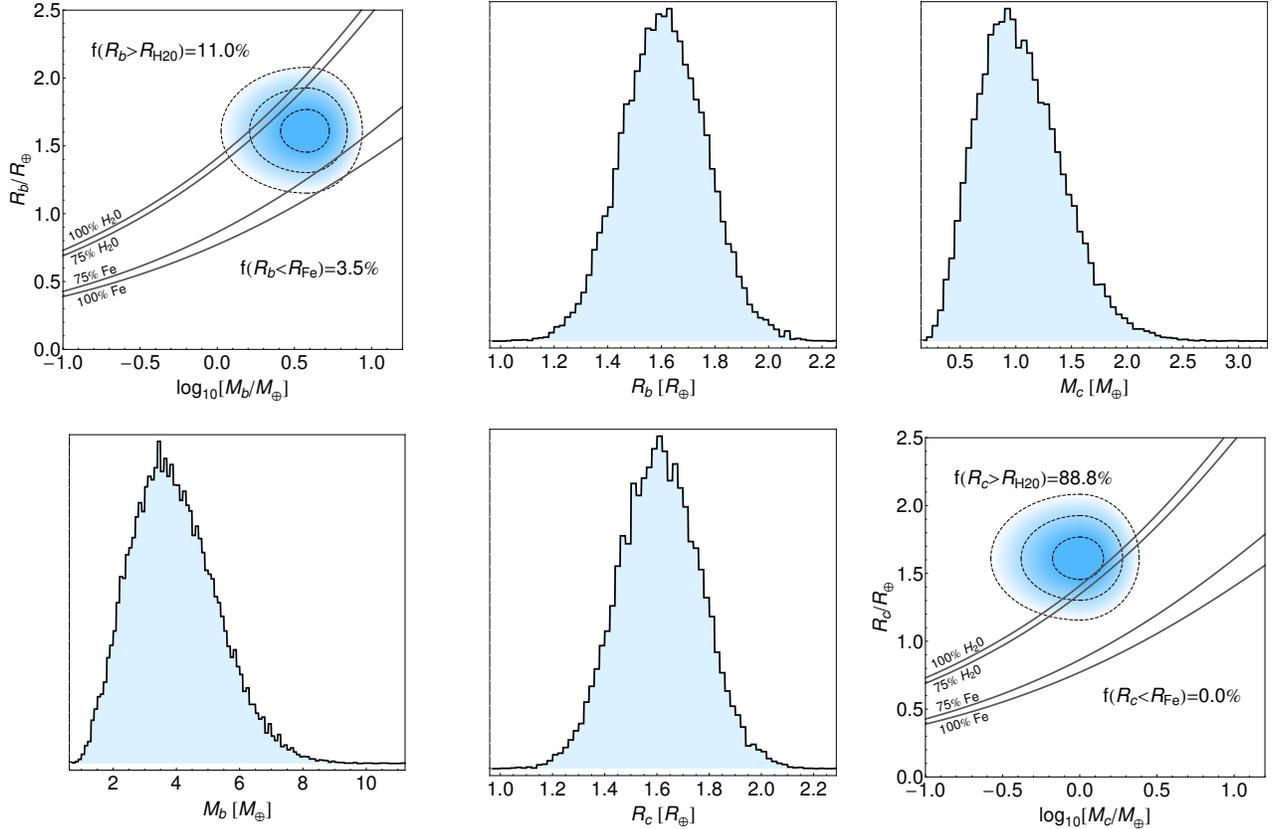}
\caption{\emph{Posterior distributions for the mass and radius of planets
KOI-314b and KOI-314c. The joint posteriors in the corners reveal that
the inner planet, b, is likely rocky, whereas the Earth-mass outer planet, c, 
likely maintains an extended atmosphere comprising $\geq17_{-13}^{+12}$\% of the 
total radius.}} 
\label{fig:GRID0314}
\end{center}
\end{figure*}
}

Formally, our dynamical model for KOI-784.01/.02 is favored over a 
non-interacting model at $\sim$7\,$\sigma$. Despite this, we prefer to consider
these two objects a tentative detection at this time. Our caution is based on
the fact that it is difficult to actually see any structure in the observed
TTVs (see Figure~\ref{fig:TTV0784}). Further, there is an apparent lack of
power in the .01 periodogram at the expected super-period and yet considerable
power at $\sim500$\,d (see Table~\ref{tab:timing}). Our best fitting TTV model
does not reproduce any power at 500\,d, but does have its dominant power at the 
super-period. For these reasons, it is unclear whether our model is just fitting
out some non-white noise component remaining in the data. Despite this, we note
that the derived solution is apparently unique, dynamically stable and again
yields physically plausible internal compositions for both objects (see
Figure~\ref{fig:GRID0784}). Our favored solution yields significant 
eccentricity for KOI-784.01 of $e=0.182_{-0.021}^{+0.014}$, which is also 
noteworthy. We attempted to repeat the fits enforcing low eccentricities but 
this yielded an implausibly dense composition for KOI-784.01 with 99.1\% of the 
trials exceeding the mass stripping limit of \citet{marcus:2009}. We argue that 
confirmation of this signal could be made by collecting further TTVs or 
identifying an object responsible for the $\sim$500\,d periodogram peak, either 
through transits or radial velocities.

\ifthenelse{\boolean{color}}{
\begin{figure*}
\begin{center}
\includegraphics[width=16.8 cm]{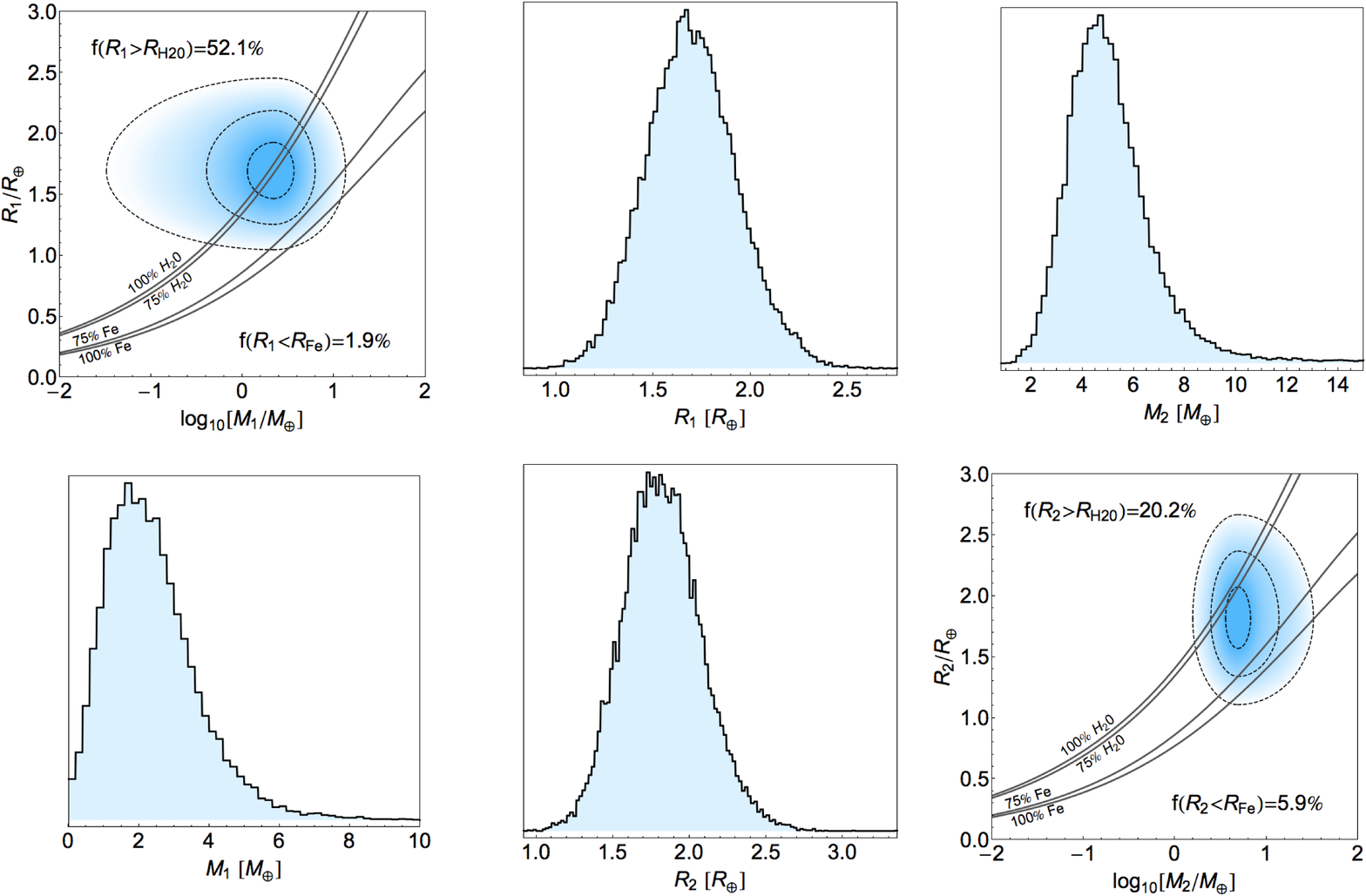}
\caption{\emph{Posterior distributions for the mass and radius of planetary
candidates KOI-784.01 and KOI-784.02. The joint posteriors in the corners show 
physically sound compositions, where the outer body (.01) has a lower density 
than the inner world (.02).}} 
\label{fig:GRID0784}
\end{center}
\end{figure*}
}{
\begin{figure*}
\begin{center}
\includegraphics[width=16.8 cm]{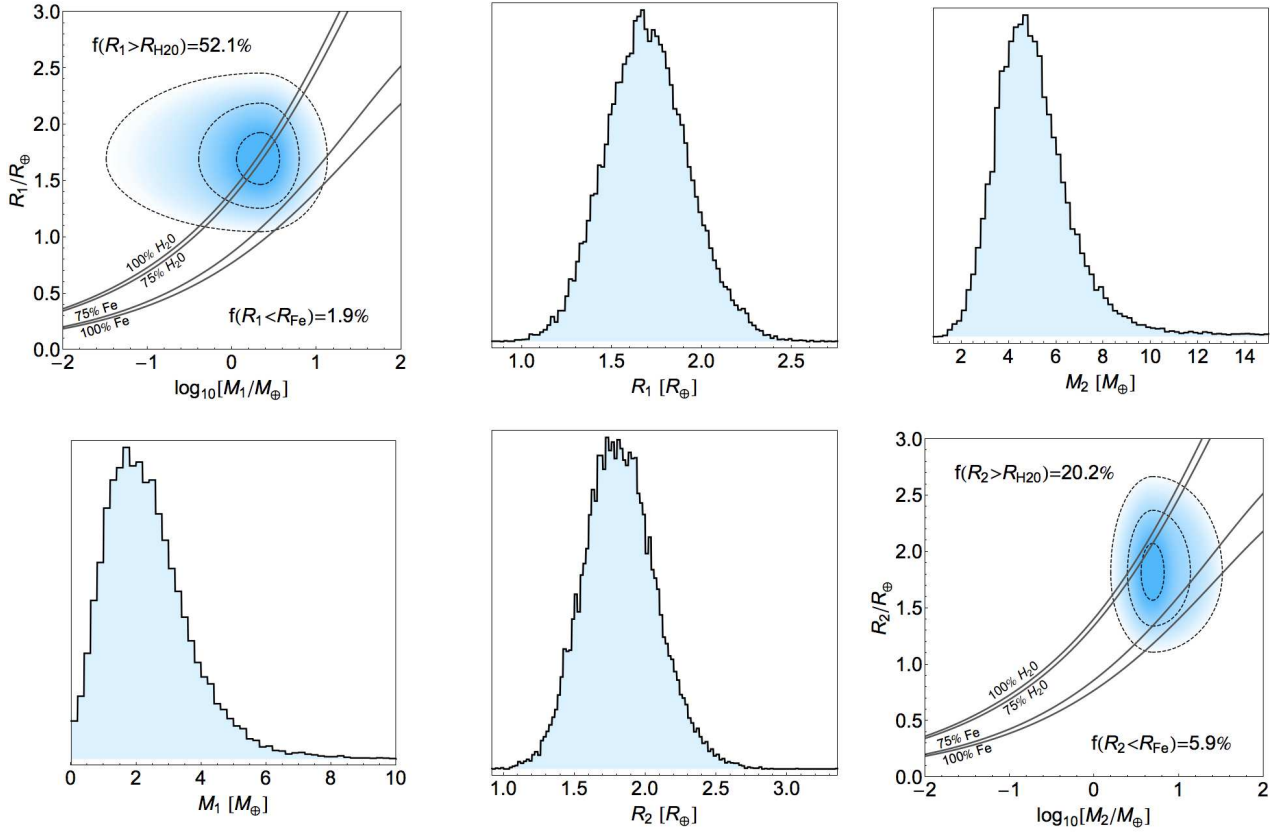}
\caption{\emph{Posterior distributions for the mass and radius of planetary
candidates KOI-784.01 and KOI-784.02. The joint posteriors in the corners show 
physically sound compositions, where the outer body (.01) has a lower density 
than the inner world (.02).}} 
\label{fig:GRID0784}
\end{center}
\end{figure*}
}

Following the strategy of \citet{koi872:2012} for KOI-872b, we adjusted the
time stamps of the detrended photometry for these two systems by offsetting
the best-fitting dynamical planet-planet model transit times. Since the
interacting planets must orbit the same star, this allows us to re-fit the
photometry with a common mean stellar density ($\rho_{\star}$) and common limb 
darkening coefficients ($q_1$ and $q_2$) using a simple planet-only model. 
Combining this information yields improved and self-consistent parameters 
for each planet. In these fits, the eccentricity and argument of periastron 
passage for the planetary candidates are fixed to the maximum a-posteriori 
values from the dynamical fits (note they are all low eccentricity). This 
combined model is referred to as $\mathcal{C}_{bc}$, where the subscript denotes 
that planets b and c were assumed to orbit the same star. Rather than attempt to 
fit a moon model through these adjusted data, we take the conservative approach 
of simply stating that we cannot place any constraints on the presence of moons 
for these KOIs due to the strong planet-planet interactions.

Our derived final parameters for the KOI-314 and KOI-784 systems are provided 
in Table~\ref{tab:dynoparams}. It is worth noting that the light curve derived 
stellar density from KOI-314b and KOI-314c ($\rho_{\star,\obs}$) is 
$\sim2$\,$\sigma$ discrepant with that expected from the stellar parameters 
derived by \citet{pineda:2013} ($\rho_{\star,\mathrm{spec}}$), with 
$(\rho_{\star,\obs}/\rho_{\star,\mathrm{spec}})=0.55_{-0.16}^{+0.23}$, whereas
the other densities appear consistent. We suggest that this is likely a result 
of the adjusted photometry used to derive $\rho_{\star,\obs}$ containing some
residual TTVs or TDVs due to either unaccounted for perturbing bodies or the
unpropagated uncertainty of the best-fitting dynamical model itself. As
discussed recently in \citet{AP:2013}, residual perturbations cause an
underestimation of $\rho_{\star,\obs}$ via the photo-timing and photo-duration
effects. Using the photo-timing expressions of \citet{AP:2013}, a residual
TTV of $\simeq$2\,minutes amplitude would be sufficient to explain the low
$(\rho_{\star,\obs}/\rho_{\star,\mathrm{spec}})$ observation. This is entirely
plausible given that the residuals of the best-fitting TTV model for
KOI-314c exhibit an r.m.s. of 4.0\,minutes. A light curve derived stellar 
density with improved accuracy and more realistic uncertainties could be derived 
using a full photodynamical model for the planet-planet interactions 
\citep{AP:2013}, however this is outside the scope of this work and it is 
unclear what insights Asterodensity Profiling (AP) could provide superior to the 
TTVs in any case for this object.

\begin{table*}
\caption{\emph{Final parameter estimates for the KOI-314 and KOI-784 systems. 
Parameters come from both a transit light curve model and a dynamical TTV
model to account for the planet-planet perturbations between KOI-314b/c and
KOI-784.01/.02. For these two pairs of objects, the transit model assumes a 
common star. Reference epochs of 770 \& 800\,BKJD$_{\mathrm{UTC}}$ are used for 
the dynamical fits of KOI-314 \& KOI-784 respectively, where 
BKJD$_{\mathrm{UTC}}=$BJD$_{\mathrm{UTC}}-2,454,833$. $^*$ = fixed quantity.
}} 
\centering 
\begin{tabular}{c c c c c c} 
\hline
KOI & 314b & 314c & 314.03 & 784.01 & 784.02 \\ [0.5ex] 
\hline
$P$ \dotfill & $13.78164_{-0.00014}^{+0.00019}$ & $23.08933_{-0.00071}^{+0.00071}$ & $10.312921_{-0.000064}^{+0.000060}$ & $19.2729_{-0.0012}^{+0.0019}$ & $10.06543_{-0.00016}^{+0.00018}$ \\
$\tau$ [BKJD$_{\mathrm{UTC}}$] \dotfill & $742.87635_{-0.00040}^{+0.00033}$ & $725.1404_{-0.0013}^{+0.0014}$ & $741.9915_{-0.0022}^{+0.0021}$ & $803.4428_{-0.0021}^{+0.0021}$ & $800.5997_{-0.0020}^{+0.0021}$ \\
$(M_P/M_{\star})\times10^{-5}$ \dotfill & $2.03_{-0.65}^{+0.76}$ & $0.53_{-0.18}^{+0.21}$ & - & $1.31_{-0.63}^{+0.84}$ & $2.89_{-0.71}^{+0.97}$ \\
$(R_P/R_{\star})$ \dotfill & $0.02730_{-0.00070}^{+0.00087}$ & $0.02731_{-0.00072}^{+0.00085}$ & $0.00753_{-0.00050}^{+0.00078}$ & $0.0322_{-0.0015}^{+0.0020}$ & $0.0345_{-0.0019}^{+0.0025}$ \\ 
$e$ \dotfill & $0.050_{-0.025}^{+0.049}$ & $0.024_{-0.016}^{+0.030}$ & - & $0.182_{-0.021}^{+0.014}$ & $0.027_{-0.018}^{+0.027}$ \\
$\omega$\,[$^{\circ}$] \dotfill & $72_{-28}^{+47}$ & $170_{-120}^{+130}$ & - & $18.4_{-5.7}^{+8.0}$ & $170_{-110}^{+130}$ \\
$\Omega$\,[$^{\circ}$] \dotfill & $270^{*}$ & $300_{-150}^{+40}$ & - & $270^{*}$ & $0_{-140}^{+140}$ \\
$\rho_{\star,\obs}$\,[g\,cm$^{-3}$] \dotfill & $2.75_{-0.47}^{+0.70}$ & $2.75_{-0.47}^{+0.70}$ & $5.3_{-1.9}^{+2.1}$ & $5.2_{-1.8}^{+1.4}$ & $5.2_{-1.8}^{+1.4}$ \\
$b$ \dotfill & $0.922_{-0.015}^{+0.011}$ & $0.767_{-0.048}^{+0.033}$ & $0.31_{-0.21}^{+0.28}$ & $0.32_{-0.22}^{+0.27}$ & $0.838_{-0.033}^{+0.046}$ \\
$q_1$ \dotfill & $0.355_{-0.051}^{+0.074}$ & $0.355_{-0.051}^{+0.074}$ & $0.79_{-0.24}^{+0.15}$ & $0.65_{-0.24}^{+0.22}$ & $0.65_{-0.24}^{+0.22}$ \\
$q_2$ \dotfill & $0.45_{-0.28}^{+0.32}$ & $0.45_{-0.28}^{+0.32}$ & $0.84_{-0.24}^{+0.12}$ & $0.58_{-0.28}^{+0.25}$ & $0.58_{-0.28}^{+0.25}$ \\
\hline
$M_P$\,[$R_{\oplus}$] \dotfill & $3.83_{-1.26}^{+1.51}$ & $1.01_{-0.34}^{+0.42}$ & - & $2.2_{-1.1}^{+1.5}$ & $4.9_{-1.3}^{+1.8}$ \\
$R_P$\,[$R_{\oplus}$] \dotfill & $1.61_{-0.15}^{+0.16}$ & $1.61_{-0.15}^{+0.16}$ & $0.446_{-0.050}^{+0.062}$ & $1.69_{-0.23}^{+0.24}$ & $1.82_{-0.25}^{+0.26}$ \\
$\rho_P$\,[g\,cm$^{-3}$] \dotfill & $5.0_{-2.0}^{+3.0}$ & $1.31_{-0.54}^{+0.82}$ & - & $2.5_{-1.4}^{+2.6}$ & $4.6_{-1.9}^{+3.7}$ \\
$R_{\mathrm{MAH}}/R_P$ \dotfill & $-0.20_{-0.18}^{+0.16}$ & $0.17_{-0.13}^{+0.12}$ & - & $0.01_{-0.23}^{+0.21}$ & $-0.27_{-0.33}^{+0.32}$ \\
$u_1$ \dotfill & $0.80_{-0.27}^{+0.19}$ & $0.80_{-0.27}^{+0.19}$ & $1.57_{-0.35}^{+0.22}$ & $1.17_{-0.35}^{+0.26}$ & $1.17_{-0.35}^{+0.26}$ \\
$u_2$ \dotfill & $-0.20_{-0.22}^{+0.32}$ & $-0.20_{-0.22}^{+0.32}$ & $-0.70_{-0.15}^{+0.27}$ & $-0.40_{-0.21}^{+0.31}$ & $-0.40_{-0.21}^{+0.31}$ \\
$S_{\mathrm{eff}}$\,[$S_{\oplus}$] \dotfill & $6.8_{-1.3}^{+1.5}$ & $3.39_{-0.65}^{+0.74}$ & $9.9_{-1.9}^{+2.1}$ & $3.34_{-0.86}^{+1.05}$ & $7.8_{-2.0}^{+2.5}$ \\ [1ex]
\hline\hline 
\end{tabular}
\label{tab:dynoparams} 
\end{table*}



\section{DISCUSSION \& CONCLUSIONS}
\label{sec:discussion}


\subsection{Discovery of an Earth-mass Planet}

Although it is not the principal goal of our work, a by-product of our 
dynamical investigations has yielded the confirmation of an Earth-mass planet, 
KOI-314c, and a super-Earth, KOI-314b. KOI-314c is the lowest mass transiting 
planet discovered to date (see Figure~\ref{fig:MR}), with a mass of 
$M_P=1.0_{-0.3}^{+0.4}$\,$M_{\oplus}$. This may be compared to the next lowest 
mass transiting object currently known, Kepler-78b \citep{sanchis:2013}, with 
$M_P=1.7\pm0.4$\,$M_{\oplus}$ \citep{howard:2013,pepe:2013} determined using the
radial velocity method. Remarkably, whereas Kepler-78b has a density 
similar to that of the Earth and thus is likely rocky, KOI-314c is 60\% larger 
than the Earth with a mean density around four times lower at 
$\rho_P = 1.3_{-0.5}^{+0.8}$\,g\,cm$^{-3}$.

Insights into the composition of this Earth-mass, but decidedly non-Earth-like,
world can be gleaned by computing the ``minimum atmospheric height'' 
($R_{\mathrm{MAH}}$), as discussed in \citet{MAH:2013}. For KOI-314c, we 
estimate $R_{\mathrm{MAH}}=0.27_{-0.22}^{+0.22}$\,$R_{\oplus}$, which would 
constitute $17_{-13}^{+12}$\% of the planet by radius fraction. The confidence
of an atmosphere being present is $89.3$\% using the technique described in
\citet{MAH:2013}. It therefore seems probable that KOI-314c is enveloped in
a gaseous light atmosphere, for which a H/He composition would be the most 
plausible candidate.

\ifthenelse{\boolean{color}}{
\begin{figure}
\begin{center}
\includegraphics[width=8.4 cm]{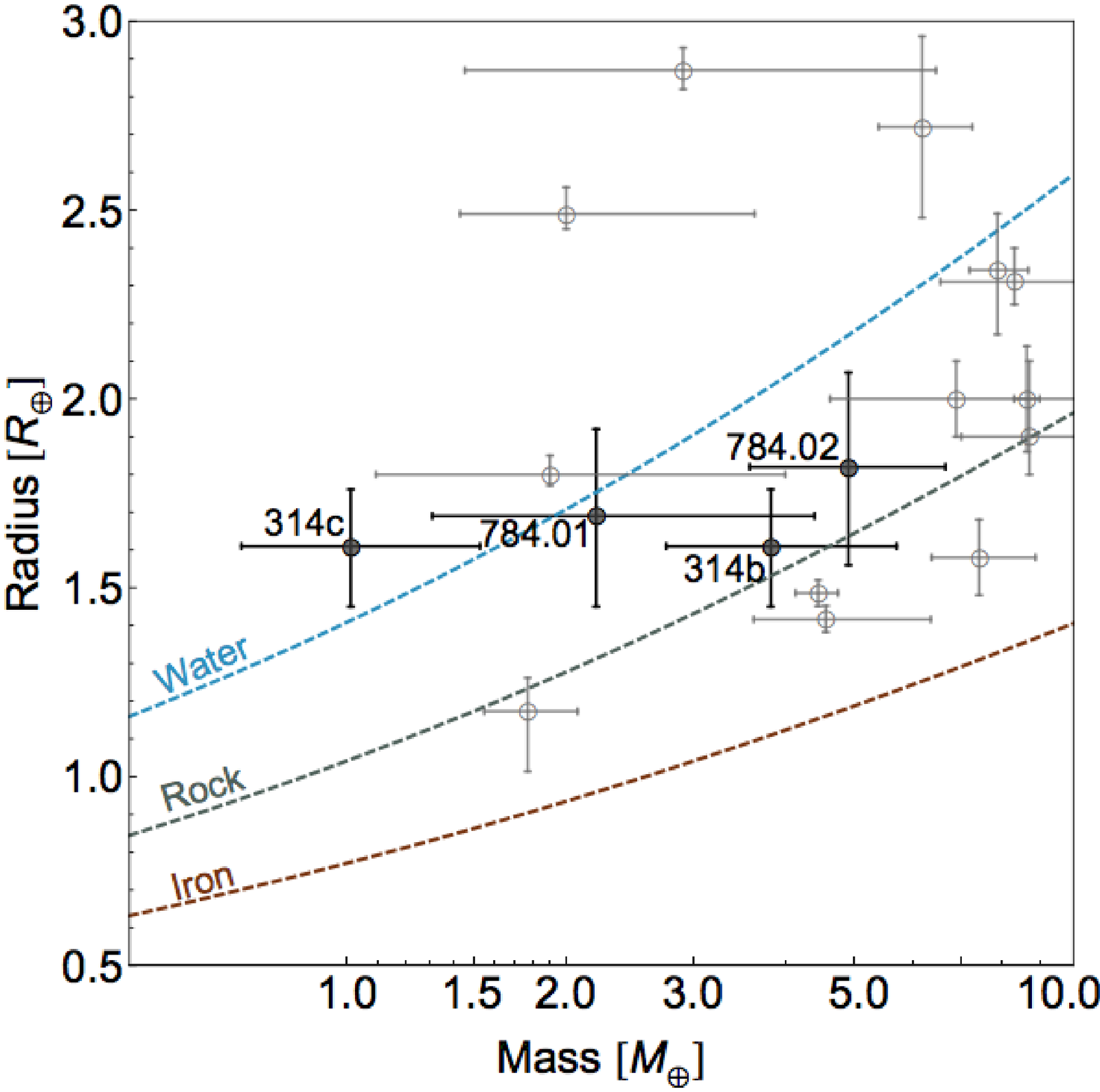}
\caption{\emph{Masses and radii of all confirmed transiting planets at the time 
of writing (gray points). The parameters for KOI-314b/c and KOI-784.01/.02 
derived in this work are shown black, which includes the lowest mass transiting
planet to date, KOI-314c. The dashed lines represent internal composition models 
from \citet{zeng:2013} assuming no atmosphere.}} 
\label{fig:MR}
\end{center}
\end{figure}
}{
\begin{figure}
\begin{center}
\includegraphics[width=8.4 cm]{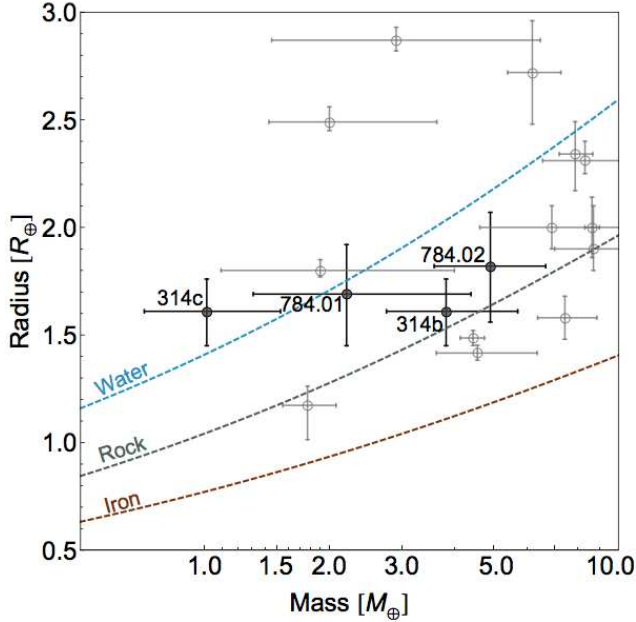}
\caption{\emph{Masses and radii of all confirmed transiting planets at the time 
of writing (gray points). The parameters for KOI-314b/c and KOI-784.01/.02 
derived in this work are shown black, which includes the lowest mass transiting
planet to date, KOI-314c. The dashed lines represent internal composition models 
from \citet{zeng:2013} assuming no atmosphere.}} 
\label{fig:MR}
\end{center}
\end{figure}
}

With a low density and thus low surface gravity of just 
$g=3.8_{-1.5}^{+2.0}$\,m\,s$^{-2}$, KOI-314c should have a considerable
scale height. Using Equation~36 of \citet{winn:2010}, we estimate that the
amplitude of the transmission spectroscopy signal may be up to 60\,ppm, which
may be compared to the transit depth of 620\,ppm i.e. $\sim$10\%. In
addition, whilst KOI-314 has a magnitude of 12.9 in \emph{Kepler's} bandpass, 
the target becomes quite bright towards the infrared with $K=9.5$ 
\citep{cutri:2003}. Thus, KOI-314c is not only the first Earth-mass transiting 
planet but also the first potentially characterizable Earth-mass planet.
Radial velocity measurements are unlikely to directly improve the mass estimate 
since we expect a semi-amplitude of $0.33_{-0.11}^{+0.13}$\,m/s due to planet c.
However, we do predict a potentially detectable $1.50_{-0.48}^{+0.56}$\,m/s due 
to planet b, whose determination may aid in refining our dynamical solution.

Both the KOI-314b/c pair and the KOI-784.01/.02 pair appear to have properties
consistent with a history of photo-evaporation. In both cases, we have one
inner planet with a density consistent with a mostly rocky world and one
outer planet with a density suggestive of a significant gaseous envelope 
(although KOI-784.02 can also be explained as being a nearly pure water world). 
This dichotomy has been seen previously with Kepler-36b/c 
\citep{carter:2012}, which \citet{lopez:2013a} attributed as a signature of
photo-evaporation. The incident bolometric fluxes for KOI-314b/c and 
KOI-784.01/.02 are modest at $\lesssim$10\,$S_{\oplus}$ (see 
Table~\ref{tab:dynoparams}) and thus less than the typical level expected to
induce photo-evaporation \citep{lopez:2013b}. However, the EUV and X-ray fluxes 
may be significantly enhanced for M-dwarf stars, such as KOI-314 and KOI-784, 
meaning it is quite plausible that photo-evaporation may be responsible for the 
observed densities.

\subsection{Exomoon Survey Results}

In this survey, we find no compelling evidence for an exomoon around any of the 
eight KOIs analyzed. Out of these eight, we find five null detections and one 
spurious photometric detection and we are able to derive robust upper limits on 
the satellite-to-planet ratio, $(M_S/M_P)$, in these cases. The other two KOIs 
are spurious dynamical detections for which we are unable to derive any upper 
limits on $(M_S/M_P)$. The final posterior distributions for $(M_S/M_P)$ are
shown in Figure~\ref{fig:Msphisto}.

\ifthenelse{\boolean{color}}{
\begin{figure*}
\begin{center}
\includegraphics[width=16.8 cm]{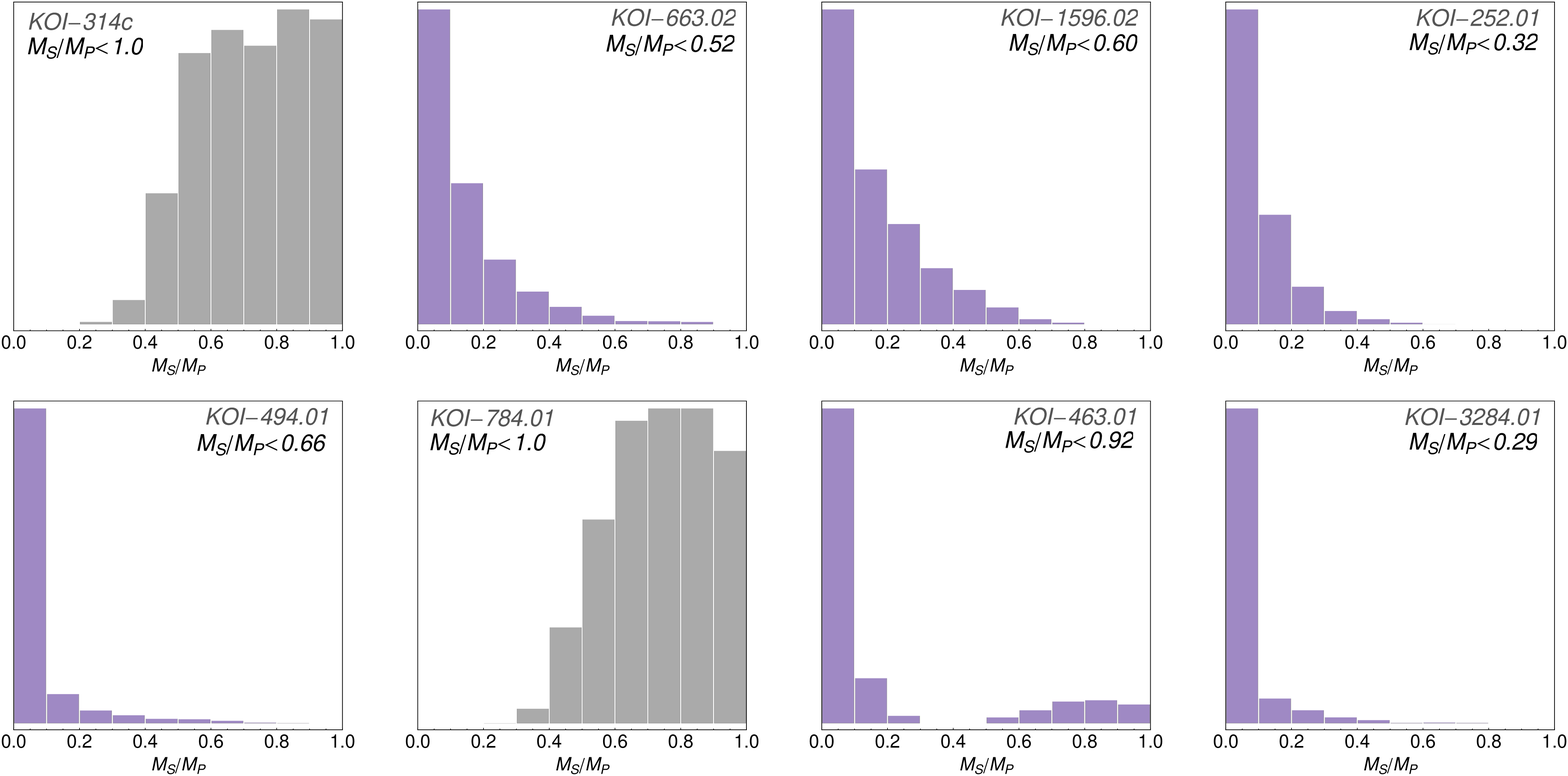}
\caption{\emph{Posterior distributions for the mass ratio between a putative
exomoon and the Kepler planetary candidates studied in this work. KOI-314c
and KOI-784.01 are dynamical spurious detections induced by planet-planet
interactions. KOI-252.01 is a photometric spurious detection and so we provide
the posterior from model $\mathcal{S}_{R0}$ to provide a corrected
estimate of the mass-ratio upper limit. The remaining five planetary candidates 
are null-detections.}} 
\label{fig:Msphisto}
\end{center}
\end{figure*}
}{
\begin{figure*}
\begin{center}
\includegraphics[width=16.8 cm]{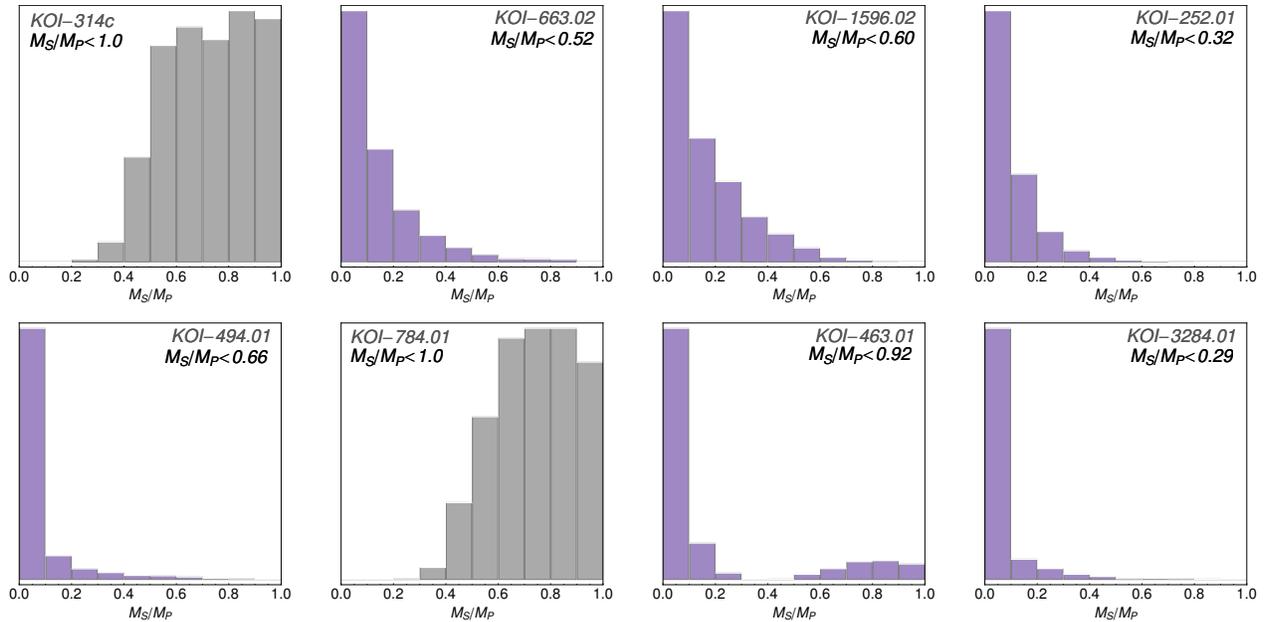}
\caption{\emph{Posterior distributions for the mass ratio between a putative
exomoon and the Kepler planetary candidates studied in this work. KOI-314c
and KOI-784.01 are dynamical spurious detections induced by planet-planet
interactions. KOI-252.01 is a photometric spurious detection and so we provide
the posterior from model $\mathcal{S}_{R0}$ to provide a corrected
estimate of the mass-ratio upper limit. The remaining five planetary candidates 
are null-detections.}} 
\label{fig:Msphisto}
\end{center}
\end{figure*}
}

Our fits only provide limits on $(M_S/M_P)$ and not $M_S$ directly. Since the
planetary masses of the six constrained cases are unknown, then $M_S$ cannot be
observationally constrained. If we assume the planetary candidates are real,
then one may invoke a mass-radius relation to provide an approximate estimate
of the typical $M_S$ values being probed in our survey. For this calculation,
we use the empirical two-component mass-radius relation recently derived by 
\citet{weiss:2013} from 63 KOIs with radii below 4\,Earth radii (appropriate for 
our sample). This yields $M_S\lesssim4.4,0.36,1.3,3.2,3.3$ \& $2.2$\,
$M_{\oplus}$ to 95\% confidence for KOI-463.01, 3284.01, 663.02, 1596.02, 
494.01 \& 252.01 respectively. This may be compared to the seven objects studied 
in survey I \citep{hek:2013}, for which we find 
$M_S\lesssim0.09,4.2,5.9,0.36,0.17,1.32$ \& $0.07$\,$M_{\oplus}$ for KOI-722.01, 
365.01, 174.01, 1472.01, 1857.01, 303.01 \& 1876.01 respectively. In general, 
survey I certainly probed down to lower $(M_S/M_P)$ ratios but the actual limits 
on $M_S$ are only modestly better for the survey I sample than this survey. 

Although survey I appears to probe down to lower limits, it is unclear exactly 
why this is the case. Although survey I defined no precise constraints on the 
stellar hosts, almost all of the stars are K/G-dwarfs with a median effective 
temperature of 5600\,K versus 3800\,K for the sample studied here. From the 
basics of exomoon detection theory, one should expect the dominant constraint to 
come from the TTV effect which scales as $\sim(M_S/M_P) a_{SB} P_P/a_{B*}$ 
\citep{sartoretti:1999,thesis:2011}. Replacing $a_{SB}$ and $a_{B*}$ with 
period-related terms via Kepler's Third Law, we therefore expected 
$(M_S/M_P)\sim\mathrm{TTV}/(p P_P)$. We find no empirical correlation between
our derived $(M_S/M_P)$ limits and this term using several reasonable metrics to 
quantify the TTV signal-to-noise. Similarly, variants of the signal-to-noise of 
the transits appear to have no clear correlation to the derived $(M_S/M_P)$ 
limits either. This may be because our sample is still relatively small and thus 
it remains difficult to identify anomalous points which skew the sample.

With this paper, the HEK project has now surveyed 17 planetary candidates
for evidence of an exomoon (\citealt{koi872:2012,hek:2013,kepler22:2013} and
this work). In all cases, we find no compelling evidence for such an object
with considerable range in the derived upper limits on the satellite-to-planet 
mass ratio down to 1\% and upper limits on the (estimated) satellite 
mass down to 0.1\,$M_{\oplus}$. As discussed in the last paragraph, it remains
unclear why such diversity exists in our carefully selected sample. However,
the best case sensitivities match the predictions made in \citet{kipping:2009},
supporting the hypothesis that \emph{Kepler} is capable of detecting sub-Earth
mass moons. At this stage, with just 17 objects surveyed, the sample size is
too small to deduce any meaningful occurrence rate statistics although it would
seem that large moons are at least not ubiquitous. In future work, we will 
expand the sample with the ultimate goal of providing a measurement of
the occurence of large moons around viable planet hosts, $\eta_{\leftmoon}$.


\acknowledgements
\section*{Acknowledgements}

We would like to thank Ren\'e Heller for his thoughtful review which improved 
the quality of our manuscript.
This work made use of the Michael Dodds Computing Facility.
DMK is funded by the NASA Carl Sagan Fellowships. JH and GB acknowledge partial 
support from NSF grant AST-1108686 and NASA grant NNX12AH91H. DN acknowledges 
support from NSF AST-1008890.
We offer our thanks and praise to the extraordinary scientists, engineers
and individuals who have made the \emph{Kepler Mission} possible. Without
their continued efforts and contribution, our project would not be possible.
%


\section*{Appendix:Planet-planet Dynamical Fits}
\label{app:TTV}


We here provide details on how the dynamical planet-planet fits to the TTVs and
TDVs were executed. In all cases, the regressions were executed using \multi\ 
\citep{feroz:2008,feroz:2009}, which is able to fully explore complex and
multi-modal parameter space and report on the uniqueness and relative 
significances of any modes identified. The dynamical model called by \multi\ 
considers the planets placed in general orbits, which are numerically integrated 
forward in time using a code based on a symplectic $N$-body integrator known as 
{\tt swift\_{}mvs} \citep{levison:1994}. {\tt swift\_{}mvs} is an efficient 
implementation of the second-order map developed by \citet{wisdom:1991} and
in practice we apply a symplectic corrector to improve the integrator's 
accuracy, as discussed in \citet{wisdom:1996}.

Our model computes the mid-transit times and durations by interpolation as 
described in \citet{koi142:2013}. With an integration time step of 1/20 of the 
orbital period, the typical precision is better than a few seconds, which is 
better than needed because the measurements generally have $>$1\,minute errors. 
We use \multi\ with a Gaussian likelihood function on the transit times and
durations, using 4000 live points and a target efficiency of 0.1, to fully
explore the parameter space and locate plausible solutions.

The low radius of KOI-314.03 coupled with the lack of period commensurability
meant that TTVs from this object would be $\lesssim$\,seconds in amplitude and
thus would be undetectable with our measurements. For both KOI-314 and KOI-784
then, we only consider dynamical two-planet fits. Accordingly, the dynamical 
model has 14 parameters: the mass ratios $(M_1/M_{\star})$ \& $(M_2/M_{\star})$, 
the orbital periods $P_1$ \& $P_2$, the time for each planet to evolve on an 
unperturbed orbit from $t_{\mathrm{ref}}$ to the mid-transit time of a 
selected transit ($\tau_1$ \& $\tau_2$; see \citealt{carter:2012}), 
eccentricities $e_1$ \& $e_2$, pericenter longitudes $\varpi_1$ \& $\varpi_2$, 
nodal longitude difference $\Delta\Omega=(\Omega_2-\Omega_1)$, impact parameters 
$b_1$ \& $b_2$, and stellar density $\rho_{\star}$.

We use the transit reference system \citep{koi872:2012}, where the nodal 
longitude $\Omega_1=270^\circ$ by definition, and we define the reference 
inclination, $\iota$, to be zero when the impact parameter, $b$, is zero. The 
reference time, $t_{\mathrm{ref}}$, was chosen to be close to the mid-transit 
time of a selected transit. Uniform priors were used for all parameters, except 
for $\rho_{\star}$, for which we used a Gaussian prior based on the spectoscopic 
stellar parameters (and held $M_*$ fixed at the best-fit value). Note that 
the parameters $P_1$ \& $P_2$ are the {\it osculating} periods at 
$t_{\mathrm{ref}}$ and as such they are not exactly equal to the {\it mean} 
periods inferred from the photometric analysis.


\begin{thebibliography}{99}
\bibitem[\protect\citeauthoryear{Barnes \& O'Brien}{2002}]{barnes:2002} 
Barnes, J. W. \& O'Brien, D. P., 2002, ApJ, 575, 1087
\bibitem[\protect\citeauthoryear{Bennett et al.}{2013}]{bennett:2013} Bennett,
D. P. et al., 2013, arXiv:1312.3951
\bibitem[\protect\citeauthoryear{Bonfils et al.}{2013}]{bonfils:2013} 
Bonfils, X. et al., 2013, A\&A, 549, 109
\bibitem[\protect\citeauthoryear{Carter et al.}{2012}]{carter:2012} 
Carter, J. A. et al., 2012, Science, 337, 556
\bibitem[\protect\citeauthoryear{Christiansen et al.}{2012}]{christiansen:2012} 
Christiansen, J. L. et al., 2012, PASP, 124, 1279
\bibitem[\protect\citeauthoryear{Cutri et al.}{2003}]{cutri:2003} 
Cutri, R. M. et al., 2003, VizieR Online Data Catalog, 2246, 0
\bibitem[\protect\citeauthoryear{Domingos et al.}{2006}]{domingos:2006} 
Domingos, R. C., Winter, O. C. \& Yokoyama, T., 2006, MNRAS, 373, 1227
\bibitem[\protect\citeauthoryear{Dong \& Zhu}{2013}]{dong:2013} 
Dong, S. \& Zhu, Z., 2013, ApJ, 778, 53
\bibitem[\protect\citeauthoryear{Dressing \& Charbonneau}{2013}]{dressing:2013} 
Dressing, C. D. \& Charbonneau, D., 2013, ApJ, 2013, 767, 95
\bibitem[\protect\citeauthoryear{Feroz et al.}{2008}]{feroz:2008} 
Feroz, F. \& Hobson, M. P., 2008, MNRAS, 384, 449
\bibitem[\protect\citeauthoryear{Feroz et al.}{2009}]{feroz:2009} 
Feroz, F., Hobson, M. P. \& Bridges, M., 2009, MNRAS, 398, 1601
\bibitem[\protect\citeauthoryear{Forgan \& Kipping}{2013}]{forgan:2013} 
Forgan, D. \& Kipping, D. M., 2013, MNRAS, 432, 2994
\bibitem[\protect\citeauthoryear{Fressin et al.}{2013}]{fressin:2013} 
Fressin, F. et al., 2013, ApJ, 677, 81
\bibitem[\protect\citeauthoryear{Heller}{2012}]{heller:2012} 
Heller, R., 2012, A\&A, 545, 8
\bibitem[\protect\citeauthoryear{Heller et al.}{2013}]{heller:2013} 
Heller, R. \& Barnes, R., 2013, AsBio, 13, 18
\bibitem[\protect\citeauthoryear{Howard et al.}{2010}]{howard:2010} 
Howard, A. W. et al., Science, 330, 653
\bibitem[\protect\citeauthoryear{Howard et al.}{2013}]{howard:2013} 
Howard, A. W. et al., 2013, Nature, 503, 381
\bibitem[\protect\citeauthoryear{Kipping et al.}{2009}]{kipping:2009} Kipping, 
D. M., Fossey, S. J., Campanella, G., 2009, MNRAS, 400,  398
\bibitem[\protect\citeauthoryear{Kipping}{2009a}]{kipping:2009a} Kipping, D. M., 
2009a, MNRAS, 392, 181
\bibitem[\protect\citeauthoryear{Kipping}{2009b}]{kipping:2009b} Kipping, D. M., 
2009b, MNRAS, 396, 1797
\bibitem[\protect\citeauthoryear{Kipping}{2010}]{binning:2010} 
Kipping, D. M., 2010, MNRAS, 408, 1758
\bibitem[\protect\citeauthoryear{Kipping \& Tinetti}{2010}]{kiptin:2010} 
Kipping, D. M. \& Tinetti, G., 2010, MNRAS, 407, 2589
\bibitem[\protect\citeauthoryear{Kipping}{2011a}]{luna:2011} 
Kipping, D. M., 2011a, MNRAS, 416, 689
\bibitem[\protect\citeauthoryear{Kipping}{2011b}]{thesis:2011} 
Kipping, D. M., 2011b, PhD thesis, University College London (astro-ph:1105.3189)
\bibitem[\protect\citeauthoryear{Kipping et al.}{2012}]{hek:2012} Kipping, 
D. M., Bakos, G. \'A., Buchhave, L. A., Nesvorn\'y, D. \& Schmitt, A.
2012, ApJ, 750, 115
\bibitem[\protect\citeauthoryear{Kipping}{2013a}]{LDfitting:2013} 
Kipping, D. M., 2013a, MNRAS, 435, 2152
\bibitem[\protect\citeauthoryear{Kipping}{2013b}]{AP:2013} 
Kipping, D. M., 2013b, MNRAS, submitted (arXiv:1311.1170)
\bibitem[\protect\citeauthoryear{Kipping et al.}{2013a}]{hek:2013} Kipping, 
D. M., Hartman, J., Buchhave, L. A., Schmitt, A., Nesvorn\'y, D. \& 
Bakos, G. \'A., 2013a, ApJ, 770, 101
\bibitem[\protect\citeauthoryear{Kipping et al.}{2013b}]{kepler22:2013} Kipping, 
D. M., Forgan, D., Hartman, J., Nesvorn\'y, D., Bakos, G. \'A., Schmitt, A. R. 
\& Buchhave, L. A., 2013b, ApJ, 777, 134
\bibitem[\protect\citeauthoryear{Kipping et al.}{2013c}]{MAH:2013} 
Kipping, D. M., Spiegel, D. S. \& Sasselov, D., 2013, MNRAS, 434, 1883
\bibitem[\protect\citeauthoryear{Levison \& Duncan}{1994}]{levison:1994} 
Levison, H. F. \& Duncan, M. J., 1994, Icarus, 108, 18
\bibitem[\protect\citeauthoryear{Lopez \& Fortney}{2013a}]{lopez:2013a} 
Lopez, E. D. \& Fortney, J. J., 2013a, ApJ, 776, 2
\bibitem[\protect\citeauthoryear{Lopez \& Fortney}{2013b}]{lopez:2013b} 
Lopez, E. D. \& Fortney, J. J., 2013b, ApJ, submitted (arXiv:1311:0329)
\bibitem[\protect\citeauthoryear{Lucy \& Sweeney}{1971}]{lucy:1971} 
Lucy, L. B. \& Sweeney, M. A. 1971, AJ, 76, 544
\bibitem[\protect\citeauthoryear{Mandel \& Agol}{2002}]{mandel:2002} Mandel, K. 
\& Agol, E., 2002, ApJ, 580, 171
\bibitem[\protect\citeauthoryear{Mann et al.}{2013}]{mann:2013} 
Mann, A. W., Gaidos, E. \& Ansdell, M., 2013, ApJ, 779, 188
\bibitem[\protect\citeauthoryear{Marcus et al.}{2009}]{marcus:2009} 
Marcus, R. A., Stewart, S. T., Sasselov, D. \& Hernquist, L., 2009,
ApJ, 700, 118
\bibitem[\protect\citeauthoryear{Mayor et al.}{2011}]{mayor:2011} 
Mayor, M. et al., 2011, arXiv:1109.2497
\bibitem[\protect\citeauthoryear{Muirhead et al.}{2012}]{muirhead:2012} 
Muirhead, P. S., Hamren, K., Schlawin, E., Rojas-Ayala, B., Covey, K. R., Lloyd, 
J. P., 2012, ApJ, 750, 37
\bibitem[\protect\citeauthoryear{Muirhead et al.}{2014}]{muirhead:2014} 
Muirhead, P. S. et al., 2014, ApJ, submitted
\bibitem[\protect\citeauthoryear{Nesvorn\'y et al.}{2012}]{koi872:2012} 
Nevsorn\'y, D., Kipping, D. M., Buchhave, L. A., Bakos, G. \'A., Hartman, J. \&
Schmitt, A. R., 2012, Science, 336, 1133
\bibitem[\protect\citeauthoryear{Nesvorn\'y et al.}{2013}]{koi142:2013} 
Nevsorn\'y, D., Kipping, D. M., Terrell, D., Hartman, J., Bakos, G. \'A. \&
Buchhave, L. A., 2013, ApJ, 777, 3
\bibitem[\protect\citeauthoryear{Pepe et al.}{2013}]{pepe:2013} 
Pepe, F. et al., 2013, Nature, 503, 377
\bibitem[\protect\citeauthoryear{Petigura et al.}{2013}]{petigura:2013} 
Petigura, E. A., Howard, A. W. \& Marcy, G. W., arXiv:1311.6806
\bibitem[\protect\citeauthoryear{Pineda et al.}{2013}]{pineda:2013} 
Pineda, S. J., Bottom, M. \& Johnson, J. A., 2013, ApJ, 767, 28
\bibitem[\protect\citeauthoryear{Sanchis-Ojeda et al.}{2013}]{sanchis:2013} 
Sanchis-Ojeda, R., Rappaport, S., Winn, J. N., Levine, A., Kotson, M. C., 
Latham, D. W., Buchhave, L. A., 2013, ApJ, 774, 54
\bibitem[\protect\citeauthoryear{Sartoretti \& Schneider}{1999}]{sartoretti:1999} 
Sartoretti, P. \& Schneider, J., 1999, A\&AS, 14, 550
\bibitem[\protect\citeauthoryear{Steffen et al.}{2012}]{steffen:2012} 
Steffen, J. H. et al., 2012, MNRAS, 421, 2342
\bibitem[\protect\citeauthoryear{Weiss \& Marcy}{2013}]{weiss:2013} 
Weiss, L. M. \& Marcy, G. W. 2013, ApJL, submitted (arXiv:1312.0936) 
\bibitem[\protect\citeauthoryear{Williams et al.}{1997}]{williams:1997} 
Williams, D. M., Kasting, J. F. \& Wade, R. A., 1997, Nature, 385, 234
\bibitem[\protect\citeauthoryear{Winn}{2010}]{winn:2010} Winn, J. N., 2010, 
\emph{Transits and Occultations}, EXOPLANETS, University of Arizona Press; ed: 
S. Seager
\bibitem[\protect\citeauthoryear{Wisdom \& Holman}{1991}]{wisdom:1991} 
Wisdom, J. \& Holman, M., 1991, AJ, 102, 1528
\bibitem[\protect\citeauthoryear{Wisdom et al.}{1996}]{wisdom:1996} 
Wisdom, J., Holman, M. \& Touma, J., 2006, FIC, 10, 217
\bibitem[\protect\citeauthoryear{Xie}{2013}]{xie:2013} 
Xie, J.-W., 2013, ApJS, 208, 22
\bibitem[\protect\citeauthoryear{Zeng \& Sasselov}{2013}]{zeng:2013} 
Zeng, L. \& Sasselov, D., 2013, PASP, 125, 227
\end{thebibliography}
\end{document}